\documentclass[aps,prl,reprint,showpacs,showkeys]{revtex4-1}
\usepackage{microtype}
\usepackage{amsthm}
\usepackage{amsmath}
\usepackage{amsfonts}
\usepackage[mathscr]{eucal}
\usepackage{amssymb,epsfig,setspace}
\usepackage{graphicx}
\usepackage{scalerel}
\usepackage{enumerate}
\usepackage{upgreek}
\usepackage{subfigure}
\usepackage{epsfig}
\usepackage{psfrag}

\newcommand{\sml}{\mbox{\small{(}}}
\newcommand{\smr}{\mbox{\small{)}}}

\newcommand{\tyl}{\mbox{\tiny{(}}}
\newcommand{\tyr}{\mbox{\tiny{)}}}
\newcommand{\tysl}{\mbox{\tiny{[}}}
\newcommand{\tysr}{\mbox{\tiny{]}}}

\newcommand{\hb}{\hat{\bf b}}

\newcommand{\0}{\mbox{\tiny{0}}}
\newcommand{\1}{\mbox{\tiny{1}}}
\newcommand{\2}{\mbox{\tiny{2}}}
\newcommand{\3}{\mbox{\tiny{3}}}
\newcommand{\4}{\mbox{\tiny{4}}}

\newcommand{\6}{\mbox{\tiny{6}}}
\newcommand{\7}{\mbox{\tiny{7}}}
\newcommand{\8}{\mbox{\tiny{8}}}
\newcommand{\9}{\mbox{\tiny{9}}}
\newcommand{\B}{\mbox{\tiny{B}}}

\newcommand{\M}{\mbox{\tiny{M}}}

\begin{document}
\title[Probing the eigenstates thermalization hypothesis with many-particle quantum walks on lattices]{Probing the eigenstates thermalization hypothesis with many-particle quantum walks on lattices}%
\author{Dibwe Pierrot Musumbu}%
\email{pierrotmus@gmail.com} \affiliation{Center for Theoretical Physics of the Polish Academy of Science, Al. Lotnikow 32/46, 02-668 Warsaw, Poland }%
\author{Maria Przybylska}%
\email{M.Przybylska@if.uz.zgora.pl} \affiliation{Institute of Physics, University of Zielona G\'ora, Licealna 9, 65--417 Zielona G\'ora, Poland }%
\author{Andrzej J.~Maciejewski} \email{andrzej.j.maciejewski@gmail.com}
\affiliation{Janusz Gil Institute of Astronomy, University of Zielona G\'ora, Licealna 9, 65--417 Zielona G\'ora, Poland.}%

\date{\today}%

\begin{abstract}
We simulate  dynamics of many-particle systems of  bosons and fermions  using  discrete time quantum walks on lattices.  We present a computational proof of a behavior of the simulated systems similar to the one observed in Hamiltonian dynamics during quantum thermalization. We record the time evolution of the entropy and the temperature of a specific particle configuration during the entire dynamics and observe how they relax to a state we call quantum walks thermal state. This observation is made on two types of lattices while simulating different numbers of particles walking on two grid graphs with 25 vertices.  In each case, we observe that the vertices counting statistics, the temperature of the indexed configuration and the dimension of the effective configuration Hilbert space relax simultaneously and remain relaxed for the rest of the many-particle quantum walks.
\end{abstract}
                      
\maketitle
\section{Introduction}
Many-particle quantum walks on graphs are relatively recent development of quantum computation dealing  with the description of the dynamics of particles in the discrete spaces modeled by a graph.
Quantum walks in general  have many applications in quantum information theory and they can describe various  physical phenomena in a wide range of such domains as relativistic physics, quantum chaos, solid state physics, quantum optics, quantum matter, condensed matter physics etc... Recent developments of experimental techniques have enabled the laboratory realization of quantum walks in many different physical systems \cite{Peruzzo2010,Karski2009,Zahringer2010}. However extensive studies have almost exclusively focussed on single particle quantum walks which do not contain the full extend of non-classical effects offered by quantum theory. Single particle quantum walks can be effectively simulated by classical systems, such as the classical light scattering in optical networks. Many-particle quantum walks are potential candidates for the development of new technics for simulations of many-body systems and help to understand various aspects of many-body quantum physics in systems of strongly correlated particles.

The studies of thermalization in quantum systems present the challenge of reconciling the time reversibility of microscopic dynamics and apparent irreversibility of the laws of thermodynamics. The most common approach in these studies is based on Hamiltonian dynamics  \cite{Rigol0,Huse2014,Gemmer2014,Srednicki2012}.  In this paper we study discrete many-particle quantum walks. The time evolution of such systems is not unitary. Anyway, our numerical experiments allow us to observe phenomenon similar to  quantum thermalization. To relate the analogy of our observation to the quantum thermalization in Hamiltonian evolution, we will use the concept of configuration and name our observed behavior quantum walks thermalization. The notion of configuration appears in many-body physics with cold atoms \cite{Immanuel01,Lewenstein:2012} where it represents the way of distributing cold atoms on optical lattices.  Experimentally a finite number of atoms are controlled using cavities representing the optical potential formed by a number of orthogonal standing waves produced by laser light beams at the center of each cavity.  Each cavity provides a position vector to a group of atoms and the optical lattice has the structure of a graph. If we associate a coordinate to each cavity on the optical lattice, distributing the $N$ indistinguishable particles on the optical lattice is similar to the problem of distributing $N$ identical balls in a finite number of buckets. In such a case each distribution of the $N$ particles on the optical lattice is an eigenstate of the Hamiltonian of the system of $N$ particles on this lattice. We use a configuration \cite{Immanuel01,Bloch2012} to implement many-partcle quantum walks. The set of all possible configurations spans the configurations Hilbert space \cite{Pierrot:01}.  

Considering the notion of thermalization in quantum systems, we  start with its short classical overview. For classical many-particle systems thermalization means convergence of mean values distributions  of physical observables such as momentum, energy etc...toward the Boltzmann distribution, independently of their initial conditions. At the classical level the mechanism responsible for thermalization is the chaos in phase space. During its evolution, the state of the system described by a point in phase space moves in an ergodic way and loses its dependence on the initial conditions. When we try to translate this thermalization mechanism to quantum framework, we encounter severe difficulties because quantum evolution is not  usually formulated in the phase space and the notion of quantum chaos is less understandable than for classical system. The Sch\"odinger equation is linear and there is no dynamical chaos in the classical sense. A big step toward understanding thermalization in quantum systems was made through the contributions in \cite{Deutsch,Srednicki} where the idea of eigenstate thermalization hypothesis (ETH) was introduced. Applications of the ETH has provided some interesting results whereas ETH remains still unproved hypothesis in spite of fact that it has been extensively studied. The quantum thermalization is more subtle than the thermalization in classical system \cite{Rigol0,Mondaini2016}. 

Mathematically quantum thermalization emerges when the time-average of an observable $\hat {O}$ converges toward the expectation value of the thermodynamical ensemble given by $\hat{\rho}_{eq}$ \cite{Srednicki2012,LangenPhd}
\vspace*{-0.2cm}
\begin{equation}
\label{0.0}
\bar {O}=\lim_{t\rightarrow\infty}\frac{1}{t}\int_{t}\langle\Psi_{r}|\hat {O}|\Psi_{r}\rangle=\frac{\operatorname{Tr}(\hat{\rho}_{eq}\hat {O})}{\operatorname{Tr}(\hat{\rho}_{eq})},
\end{equation}
where the last equality is an hypothesis.
Let us consider a system with one conserved macroscopic observable such as energy or number of particles. We assume that it is initially prepared in a pure state $\hat{\rho}_{\tyl{t=0}\tyr}=|\varphi\rangle\langle\varphi|$ or a set of coherent pure states 
\vspace*{-0.2cm}
\begin{equation}
\label{0.1}
\hat{\rho}_{\tyl{t=0}\tyr}=\sum_{\ell}a_{\ell}^*a_{\ell}|\varphi_{\ell}\rangle\langle\varphi_{\ell}|,
\vspace*{-0.25cm}
\end{equation}
where the $|\varphi_{\ell}\rangle$ are pure states and $a_{\ell}$ the coherence amplitudes.
The mean value of an observable $\hat {O}$ is defined by
\vspace*{-0.5cm}
 \begin{equation}
\label{0.2}
\langle\hat {O}\rangle =\sum_{\ell_{\1},\ell_{\2}}a_{\ell_{\1}}^*a_{\ell_{\2}}e^{i\tyl E_{\ell_{\1}}-E_{\ell_{\2}}\tyr t}|\varphi_{\ell_{\2}}\rangle\langle\varphi_{\ell_{2}}|\hat {O}|\varphi_{\ell_{\1}}\rangle\langle\varphi_{\ell_{1}}|.
\vspace*{-0.25cm}
\end{equation}
Infinite time average of this mean value $\bar {O}$ causes vanishing of the off diagonal terms and Eq. ~\eqref{0.2} gives 
\vspace*{-0.25cm}
\begin{equation}
\label{0.3}
\bar {O} =\sum_{\ell}a_{\ell}^*a_{\ell}|\varphi_{\ell}\rangle\langle\varphi_{\ell}|\hat {O}|\varphi_{\ell}\rangle\langle\varphi_{\ell}|.
\vspace*{-0.25cm}
\end{equation}
Physically we say that the system dynamically evolves in time to a state where the coherence between particular states is destroyed during its relaxation to a thermal state. The thermal state in this case is defined as
\vspace*{-0.25cm}
 \begin{equation}
\label{0.4}
\hat{\rho}_{\tyl{t}\tyr}=\sum_{\ell}P_{\ell}|\varphi_{\ell}\rangle\langle\varphi_{\ell}|,
\vspace*{-0.4cm}
\end{equation}
where $P_{\ell} \sim e^{-\beta E_{\ell}}$, $\beta=\frac{1}{k_{\B}T}$, $E_{\ell}$'s are the eigen-energies, $k_{\B}$ the Boltzmann constant and $T$ the temperature. The phenomenon of passage from an initial state  (a pure state or a set of coherent pure states) to a final thermal state is called thermalization. This phenomenon is usually explained by the ETH which states that all eigenstates of isolated quantum bounded systems thermalize individually. It means that each eigenstate contains a thermal state. We need to indicate that in both cases; starting with one pure state or a set of coherent pure states \cite{Rigol01,Huse:01,Huse:02}, the components of the initial state remain present all the time in the dynamics of the system.

The main aim of this article is to use many-particle quantum walks on $M^{\2}$ vertices lattice and implement numerical experiments which present evidences of a behavior similar to the ETH in quantum thermalization. We will apply the framework of many-particle quantum walks on graphs \cite{Pierrot:01} to two lattices with the same type of quantum walkers starting in the same initial pure state and simulate the time evolution of the temperature of this state, the vertices counting statistics and the dimension of the effective configurations Hilbert space. The effective configurations Hilbert space $\mathcal{H}_e$ is a subspace of the configurations Hilbert space $\mathcal{H}_V$ containing all the configurations with nonzero amplitudes forming the state of system.  We will study how this behavior similar to quantum  thermalization emerges during the time evolution of a configuration. We also show how this behavior which we call quantum walks thermalization is correlated to the reaching the limit value of the dimension of the effective configurations Hilbert space and the change of regime in the vertices counting statistics. 
\section{Lattice many-particle state}

We are interested in quantum closed systems consisting of $N$ indistinguishable particles on a $M^{\2}$ vertices lattice.  We will consider fermions as well as bosons. In our description of particles on lattice we will use field operators introduced by Jordan and Wigner \cite{Wigner01}. The position field creation operator $\widehat{\mathbf{\Uppsi}}^{\dagger}{\sml x_{\alpha}\smr}$ creates
a particle at vertex  ${\alpha}$:
\vspace*{-0.2cm}
\begin{equation}
 \label{2.1}
 \widehat{\mathbf{\Uppsi}}^{\dagger}{\sml x_{\alpha}\smr}|\Omega\rangle=|x_{\alpha}\rangle
\end{equation}
by acting on the vacuum state $|\Omega\rangle$ and its Hermitian conjugate called the position field annihilation operator $\widehat{\mathbf{\Uppsi}}{\sml x_{\alpha}\smr}$  destroys a particle at vertex $\alpha$: 
\vspace*{-0.2cm}
\begin{equation}
 \label{2.2}
 \widehat{\mathbf{\Uppsi}}{\sml x_{\alpha}\smr}|{x_{\alpha}}\rangle=|\Omega\rangle.
\end{equation}  
 The field creation $\widehat{\mathbf{\Uppsi}}^{\dagger}{\sml x_{\2}\smr}$ and the annihilation $\widehat{\mathbf{\Uppsi}}{\sml x_{\1}\smr}$
 operators satisfy the commutation relation and anti-commutation relations,
 \vspace*{-0.2cm}
\begin{eqnarray}
 \label{2.3}
 [\widehat{\mathbf{\Uppsi}}{\sml{x}_{\1}\smr},\,\widehat{\mathbf{\Uppsi}}^{\dagger}{\sml{x}_{\2}\smr}]_{\mp}&=&\delta(x_{\1}-x_{\2}),\\
  \vspace*{-0.3cm}
 \label{2.4}
 [\widehat{\mathbf{\Uppsi}}{\sml{x}_{\1}\smr},\,\widehat{\mathbf{\Uppsi}}{\sml{x}_{\2}\smr}]_{\mp}&=&[\widehat{\mathbf{\Uppsi}}^{\dagger}{\sml{x}_{\1}\smr},\,\widehat{\mathbf{\Uppsi}}^{\dagger}{\sml{x}_{\2}\smr}]_{\mp}=0,
\end{eqnarray}
for bosons and fermions, respectively. Here $[ \cdot,\cdot ]_{-}$ denotes the commutator and $[  \cdot,\cdot ]_{+}$ is the anti-commutator.
We define the vertex occupation number operator $\hat{\bf n}_{\alpha}=\widehat{\mathbf{\Uppsi}}^{\dagger}{\sml x_{\alpha}\smr}\widehat{\mathbf{\Uppsi}}{\sml x_{\alpha}\smr}$. The quantum state of a vertex $\alpha$ with $n$ particles can be written as 
 \vspace*{-0.2cm}
\begin{equation}
 \label{2.5}
 |n_{\alpha}\rangle=\frac{(\widehat{\mathbf{\Uppsi}}^{\dagger}{\sml x_{\alpha}\smr})^{n}}{\sqrt{n!}}|\Omega\rangle.
\end{equation} 
Moreover, it should be mentioned in Eq. \eqref{2.5} that the vertex occupation number $n$ is restricted to $n=0$ or $n=1$ for fermions.
The state $|n_{\alpha}\rangle$ is an eigenstate of the vertex occupation number
 \vspace*{-0.2cm} 
\begin{equation}
 \label{2.6}
\hat{\bf n}_{\alpha} |n_{\alpha}\rangle={n}_{\alpha} |n_{\alpha}\rangle.
\end{equation} 
In this work we consider non-interacting particles.
The position field creation and annihilation operators are connected to the multimode Fock creation and annihilation operators by the following basis transformations:
\begin{equation}
 \label{2.8}
  \hspace*{-0.2cm} {\hb}_{\upeta}^{\dagger}=\sum_{\alpha}\frac{e^{i\varphi_{\upeta}{x}_{\alpha}}}{\sqrt{N}}\widehat{\mathbf{\Uppsi}}^{\dagger}{\sml {x}_{\alpha}\smr},\quad{\hb}_{\upeta}=\sum_{\alpha}\frac{e^{-i\varphi_{\upeta}{x}_{\alpha}}}{\sqrt{N}}\widehat{\mathbf{\Uppsi}}{\sml {x}_{\alpha}\smr}.
\end{equation}
These transformations conserve the commutation (anti-commutation) rules.
\begin{figure}[htb]
    \centering
       \subfigure[Particles on lattice {I}]
    {
        \includegraphics[width=3cm]{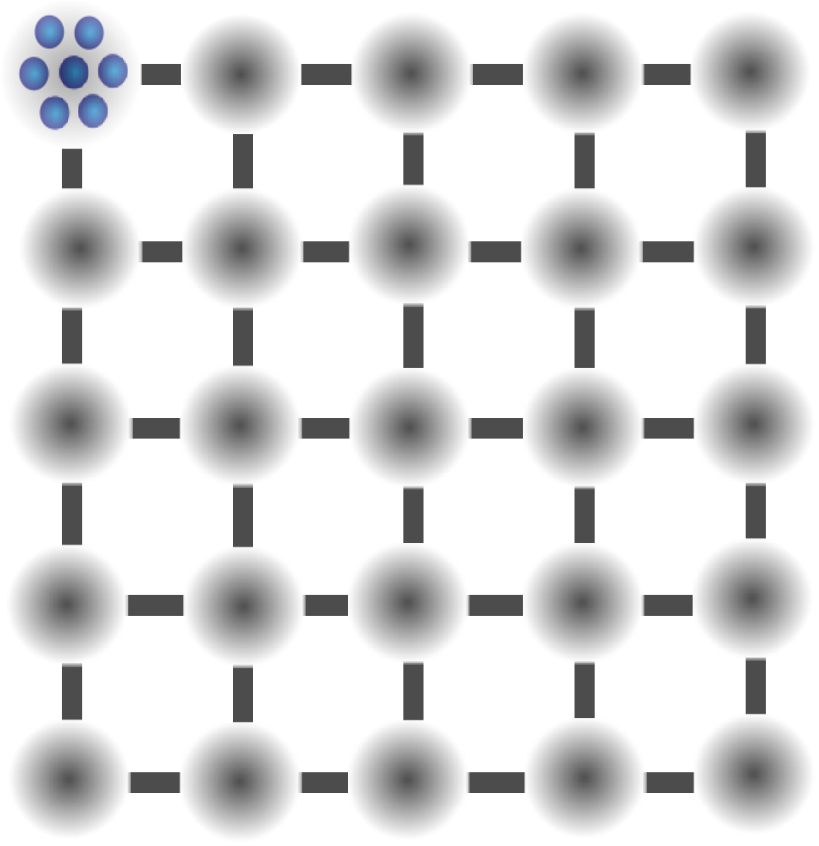}
        \label{fig:starting3}
    }
    \subfigure[Particles on lattice {II}]
    {
        \includegraphics[width=3cm]{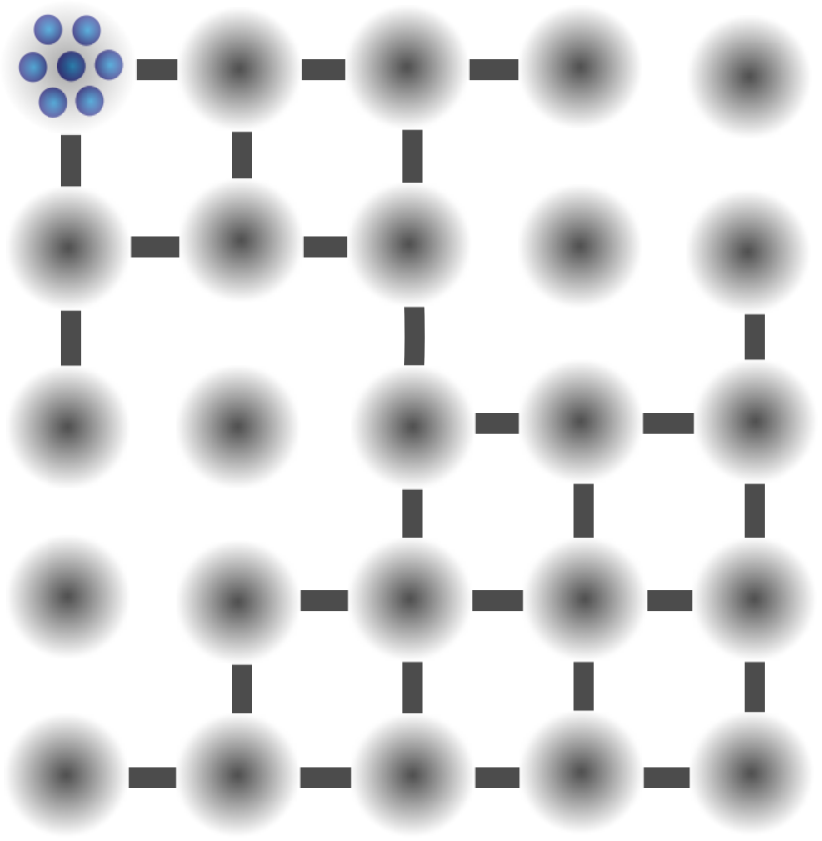}
        \label{fig:starting4}
    }
     \vspace*{-0.25cm}
    \caption{Initial configurations of $N$ particles initially prepared for quantum walks on lattices. The lattice {I} is a fully connected ${5\times5}$ grid graph and the lattice {II} is a  ${5\times5}$ partially connected grid graph with four isolated vertices.}
    \label{fig:starting}
\end{figure}
We consider two types of  lattices given in Figs.~\ref{fig:starting}. We label the vertices using numbers $\alpha=\#{\mbox{column}}+(\#{\mbox{row}}-1)\times\#{\mbox{column}}$. In lattice I each vertex is connected to all neighboring vertices i.e. we have the grid graph, see Fig.~\ref{fig:starting3}. Lattice {II} consists of two separated sub-lattices with 8 and 13 vertices respectively that are joined just by one central vertex and additionally we have four completely separated vertices, see Figs.~\ref{fig:starting}. This lattice that we call  lattice II has been previously used in \cite{Rigol0} to study the thermalization in generic isolated quantum systems. It was used as an illustration of quantum system attached to a reservoir. The reservoir is the 13 vertices sub-lattice while the proper quantum system is given by the 7 vertices sub-lattice.

We consider four systems on two types of lattices i.e. $N$ fermions and $N$ bosons on the lattice I and lattice II, respectively.  Any particular distribution of these particles on the vertices is called a configuration \cite{Immanuel01} and to describe it we use the vector  $|{\bf{n}}_{\ell}\rangle$  
\vspace*{-0.2cm}
\begin{equation}
 \label{2.9}
 |{\bf{n}}_{\ell}\rangle=\sbox0{$\begin{array}{ccc}
 {n}_{\1} & \ldots & {n}_{\M}\\
 \vdots&\ddots& \vdots\\
 {n}_{\M^{\2}-\M+1} & \ldots & {n}_{\M^{\2}} \end{array}$}
\mathopen{\resizebox{1.3\width}{\ht0}{$\Bigg|$}}
\usebox{0}
\mathclose{\resizebox{1.3\width}{\ht0}{$\Bigg\rangle$}}, 
\end{equation}
where $n_{\alpha}$ is the particle occupation number at vertex $\alpha, \alpha\in\{1,2,\ldots,\M^{\2}\}$ and $\sum_{\alpha}n_{\alpha}=N$. 
We use the tabular ket in Eq. \eqref{2.9}  for better visualization  of the notion of configuration in the case of the lattices in Figs.~\ref{fig:starting}.
The configurations Hilbert space $\mathcal{H}_{V}$ of $N$ particles on $M^{\2}$ vertices graph is spanned by vectors $|{\bf n}_{\ell}\rangle$  of the form \eqref{2.9}, where  ${\ell}$ enumerates particular configurations. Dimension of  $\mathcal{H}_V$ is
\vspace*{-0.2cm}
\begin{equation}
\label{1}
D\sml \mbox{\small{$N,M^{\2}$}}\smr=\binom{\mbox{\small{$M^{\2}+N-1$}}}{\mbox{\small{$N$}}},
\end{equation} 
see e.g. \cite{Pierrot:01}. As previously mentioned, vertices occupation numbers $n_{\alpha}$ can take any available integers number $\leq N$ for bosons and only zero or one  for fermions. For the rest of this paper we will simply use the symbol $D$ to denote the dimension of $\mathcal{H}_{V}$.

The discrete time quantum walk requires an auxiliary Hilbert space called coin's Hilbert space $\mathcal{H}_{C}$ \cite{Watrous:01}. Such a Hilbert space is spanned by the vectors specifying the directions of the edges connected to any given vertex on the lattice. The lattices used in this work are of degree four  because it is the largest number of edges connected to a vertex, see e.g.  \cite{Diestel:2012}. Therefore 
coin's Hilbert space $\mathcal{H}_{C}$ is spanned by the basis $\{|v_{\1}\rangle,|v_{\2}\rangle,|v_{\3}\rangle,|v_{\4}\rangle\}$. These four vectors are also known as coin
chiralities \cite{Watrous, Julia:02}.  

We define the Hilbert space $\mathcal{H}$ of $N$ particles on a $M^{\2}$
vertices lattice augmented with the coin chiralities as tensor product Hilbert space $\mathcal{H}=\mathcal{H}_{C}\otimes\mathcal{H}_{V}$.  Vectors $|\Psi_{r}\rangle$  of this Hilbert space are called graph many-particle states (GMP states) and determine  distributions of $N$ particles on  such lattice augmented by the coin. Normalized GMP state  $|\Psi_{r}\rangle\in\mathcal{H}$ can be written as
\vspace*{-0.2cm}
\begin{equation}
 \label{3}
|\Psi_{r}\rangle=\sum_{\ell}\sum_{i}\frac{C_{i\ell}^{r}}{\mathcal{K}_{r}}|v_{i},\,{\bf
n}_{\ell}\rangle,
\end{equation}
where
\vspace*{-0.2cm}
\begin{equation}
 \label{3.1}
[\mathcal{K}_{r}]^2=\langle\Psi_{r}|\Psi_{r}\rangle=\sum_{\ell}\sum_{i}|C_{i\ell}^{r}|^{2}
\end{equation}
and $\mathcal{K}_{r}$ is the normalization constant. The index $r$ indicates the time step. Suppose that we initialize the quantum walks by preparing the system in a single configuration of the configurations Hilbert space. Then the summation in formula \eqref{3} over all configurations will reduce to one term, and for the beginning of evolution $r = 0$. For example, when all the quantum walkers are initially on vertex $\alpha=1$ we have: 
\vspace*{-0.2cm}
\begin{equation}
\label{4}
|\Psi_{\0}\rangle=\sum_{i}\frac{C_{i\ell_{\0}}^{\0}}{\mathcal{K}_{\0}}|v_{i},\,{\bf n}_{\ell_{\0}}\rangle,
\end{equation}
where 
\begin{equation}
\label{4.1}
|v_{i},\,{\bf n}_{\ell_{\0}}\rangle=|v_{i}\rangle\sbox0{$\begin{array}{ccccc}
 N & 0 & 0 & 0 & 0\\
 0 & 0 & 0 & 0 & 0\\
 0 & 0 & 0 & 0 & 0\\
 0 & 0 & 0 & 0 & 0\\
 0 & 0 & 0 & 0 & 0\end{array}$}
\mathopen{\resizebox{1.2\width}{\ht0}{$\Bigg|$}}
\usebox{0}
\mathclose{\resizebox{1.2\width}{\ht0}{$\Bigg\rangle$}}
\end{equation}
 is the configuration with all the particles at vertex $\alpha=1$. 

\section{The conditional shift operator}

In our previous work \cite{Pierrot:01} we described in detail the construction of the conditional shift operator that generate quantum walks dynamics on graph.  Here we consider two dimensional lattices where most of the vertices have four
 incident edges. Such lattices are graphs of degree four. The presence of edges between particular vertices is characterized by the adjacency matrix that for $M^{\2}$ verteices graph is given by $M^{\2}\times{M}^{\2}$ square matrix with entries
 \begin{eqnarray}
A_{\mu\nu} &=&  \begin{cases}
1  &   ~~  \text{when $\mu$ and $\nu$ are connected,} \\
0  &   ~~ \text{when $\mu$ and $\nu$ are not connected.} 
\end{cases}\nonumber
\end{eqnarray}
 For considered lattices the adjacency matrix splits into four components that are two pairs of mutually transposed components  ${\bf
 A}_{\mbox{\tiny{L}}}$ and ${\bf A}_{\mbox{\tiny{L}}}^{\mbox{\tiny{T}}}$ as well as ${\bf A}_{\mbox{\tiny{D}}}$ and ${\bf A}_{\mbox{\tiny{D}}}^{\mbox{\tiny{T}}}$. 
 \vspace*{-0.2cm}
  \begin{equation}
  \label{5}
  {\bf A}=\sum_{k=1}^{\4}{\bf A}^{k}={\bf A}_{\mbox{\tiny{L}}}+{\bf A}_{\mbox{\tiny{L}}}^{\mbox{\tiny{T}}}+{\bf A}_{\mbox{\tiny{D}}}+{\bf
  A}_{\mbox{\tiny{D}}}^{\mbox{\tiny{T}}}.
\end{equation}
The component ${\bf A}_{\mbox{\tiny{L}}}$ of the adjacency matrix ${\bf A}$
represents all the leftward oriented connections between vertices.
In other words two vertices $\mu$ and $\nu$ are left wise connected if and only
if $\mu-\nu=1$. The component ${\bf A}_{\mbox{\tiny{D}}}$ of the adjacency
matrix ${\bf A}$ represents all the downward oriented connectivities i.e.
two vertices $\mu$ and $\nu$ are down wise connected if and only
if $\mu-\nu=M$. 

We use the same coin for all particles and we chose $4$-dimensional discrete Fourier transform coin,
see e.g \cite{Barry02}. The coin operations in this case are performed using a four sides dice and each side of the dice represents the action of a specific component of the adjacency matrix 
\begin{equation}
\label{7}
H_{4}=\frac{1}{2}\sum_{j,k=1}^{\4}e^{i\mbox{\tiny{$\frac{2\pi}{4}$}}\tyl k-1\tyr\tysl j-1\tysr}|v_{k}\rangle\langle v_{j}|.
\end{equation}
The coin tossing operation gives:
\begin{eqnarray}
\label{7.1}
H_{4}|v_{l}\rangle&=&\frac{1}{2}\sum_{j,k=1}^{\4}e^{i\mbox{\tiny{$\frac{2\pi}{4}$}}\tyl k-1\tyr\tysl j-1\tysr}|v_{k}\rangle\langle v_{j}|v_{l}\rangle\nonumber\\
&=&\frac{1}{2}\sum_{k=1}^{\4}e^{i\mbox{\tiny{$\frac{2\pi}{4}$}}\tyl k-1\tyr\tysl l-1\tysr}|v_{k}\rangle\delta_{jl},
\end{eqnarray}
and explicit formulas are the following
\begin{eqnarray}
\label{7.2}
H_{4}|v_{1}\rangle&=&\frac{1}{2}\big(|v_{\1}\rangle+|v_{\2}\rangle+|v_{\3}\rangle+|v_{\4}\rangle\big),\\
\label{7.3}
H_{4}|v_{2}\rangle&=&\frac{1}{2}\big(|v_{\1}\rangle+e^{i\mbox{\tiny{$\frac{\pi}{2}$}}}|v_{\2}\rangle+e^{i\pi}|v_{\3}\rangle+e^{i\mbox{\tiny{$\frac{3\pi}{2}$}}}|v_{\4}\rangle\big),\\
\label{7.3}
H_{4}|v_{3}\rangle&=&\frac{1}{2}\big(|v_{\1}\rangle+e^{i\pi}|v_{\2}\rangle+e^{i\2\pi}|v_{\3}\rangle+e^{i\3\pi}|v_{\4}\rangle\big),\\
\label{7.3}
H_{4}|v_{4}\rangle&=&\frac{1}{2}\big(|v_{\1}\rangle+e^{i\mbox{\tiny{$\frac{\3\pi}{2}$}}}|v_{\2}\rangle+e^{i\3\pi}|v_{\3}\rangle+e^{i\mbox{\tiny{$\frac{\9\pi}{2}$}}}|v_{\4}\rangle\big).
\end{eqnarray}
The dynamics of the quantum walks is generated by the shift operator
 that shift quantum walkers between adjacent vertices
 \begin{equation}
\label{7.4}
{\bf S}=\sum_{\mu,\nu}\sum_{k}\widehat{\mathbf{\Uppsi}}^{\dagger}{\sml {x}_{\nu}\smr}A_{\nu\mu}^{k}\widehat{\mathbf{\Uppsi}}{\sml {x}_{\mu}\smr}\otimes|v_k\rangle\langle v_k|.
\end{equation}
It is constructed as the product of the adjacency matrix that describes the connectivity of a given vertex with its neighbors and the field  annihilation and creation field operators. The field operators  annihilate a particle at vertex $\mu$ and create a particle at vertex $\nu$ shifting the particle from $\mu$ to $\nu$. Finally we define the conditional shift operator as $\widehat{\mathcal{S}}=({\bf S}\otimes{\bf I})H_{4}$ where ${\bf I}$ is the identity operator, for more detail see \cite{PierrotPhd}. Thus it can simply be written as
\vspace*{-0.2cm}
\begin{equation}
 \label{8}
  \widehat{\mathcal{S}}=\sum_{\mu,\nu}\sum_{k}\widehat{\mathbf{\Uppsi}}^{\dagger}{\sml {x}_{\nu}\smr}A_{\nu\mu}^{k}\widehat{\mathbf{\Uppsi}}{\sml {x}_{\mu}\smr}\otimes\sum_{j}\frac{e^{i\mbox{\tiny{$\frac{2\pi}{4}$}}\tyl k-1\tyr\tysl j-1\tysr}}{2}|v_j\rangle\langle v_k|.
\end{equation}
Here $A_{\nu\mu}^{k}$ indicates the action of the permutation group of the graph on its adjacency matrix \cite{Pierrot:01}. Such an operation associates an element $A_{\nu\mu}$ with the chirality $k$ of the coins.

\section{Implementation of many-particle quantum walks}

We implement the first step $|\Psi_{\1}\rangle=\widehat{\mathcal{S}}|\Psi_{\0}\rangle$ of many-particle quantum walks on a $M^{\2}$ vertices lattice explicitly
\vspace*{-0.1cm}
\begin{eqnarray}
\label{9}
  |\Psi_{\1}\rangle&=&\widehat{\mathcal{S}}|v_{i},\,{\bf n}_{\ell}\rangle=\sum_{j}\frac{C_{\2\ell_{\0}}^{\0}e^{i\mbox{\tiny{$\frac{\pi}{2}$}}\tysl j-1\tysr}}{2\mathcal{K}_{\0}}\sqrt{N}|v_{j}\,{\bf n}_{\ell_{\1}}\rangle\nonumber\\
  \vspace*{-0.2cm}
  &&\qquad\qquad+\sum_{j}\frac{C_{\3\ell_{\0}}^{\0}e^{i\pi\tysl j-1\tysr}}{2\mathcal{K}_{\0}}\sqrt{N}|v_{j}\,{\bf n}_{\ell_{\2}}\rangle,
 \vspace*{-0.3cm}
\end{eqnarray}
where  
\begin{equation}
\label{9.1}
|{\bf n}_{\ell_{\1}}\rangle=\sbox0{$\begin{array}{ccccc}
 N-1 & 1 & 0 & 0 & 0\\
 0 & 0 & 0 & 0 & 0\\
 0 & 0 & 0 & 0 & 0\\
 0 & 0 & 0 & 0 & 0\\
 0 & 0 & 0 & 0 & 0\end{array}$}
\mathopen{\resizebox{1.2\width}{\ht0}{$\Bigg|$}}
\usebox{0}
\mathclose{\resizebox{1.2\width}{\ht0}{$\Bigg\rangle$}},
\vspace*{-0.3cm}
\end{equation}
\vspace*{-0.3cm}
 and 
 \begin{equation}
 \label{9.2}
 |{\bf n}_{\ell_{\2}}\rangle=\sbox0{$\begin{array}{ccccc}
 N-1 & 0 & 0 & 0 & 0\\
 1 & 0 & 0 & 0 & 0\\
 0 & 0 & 0 & 0 & 0\\
 0 & 0 & 0 & 0 & 0\\
 0 & 0 & 0 & 0 & 0\end{array}$}
\mathopen{\resizebox{1.2\width}{\ht0}{$\Bigg|$}}
\usebox{0}
\mathclose{\resizebox{1.2\width}{\ht0}{$\Bigg\rangle$}},
 \end{equation}
 where in configuration $|{\bf n}_{\ell_2}\rangle$ one appears at vertex $\alpha=M+1$. 
\begin{figure}[htb]
    \centering
       \subfigure[configuration $\ell_{\0}$]
    {
        \includegraphics[width=2.7cm]{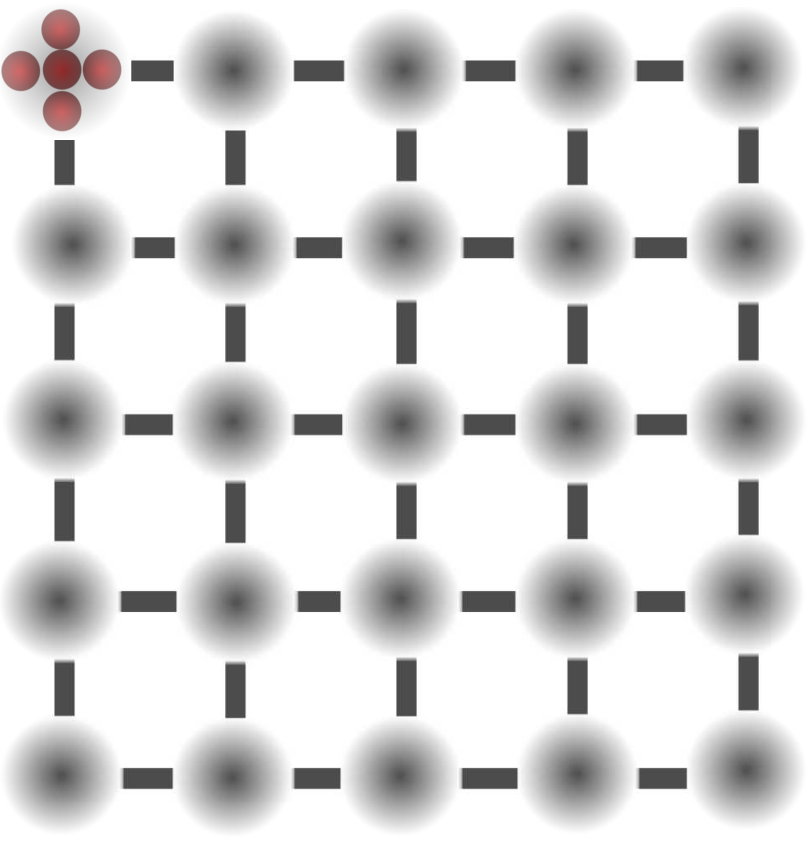}
        \label{fig:initial}
    }
    \hspace*{-0.3cm}
    \subfigure[configuration $\ell_{\1}$]
    {
        \includegraphics[width=2.7cm]{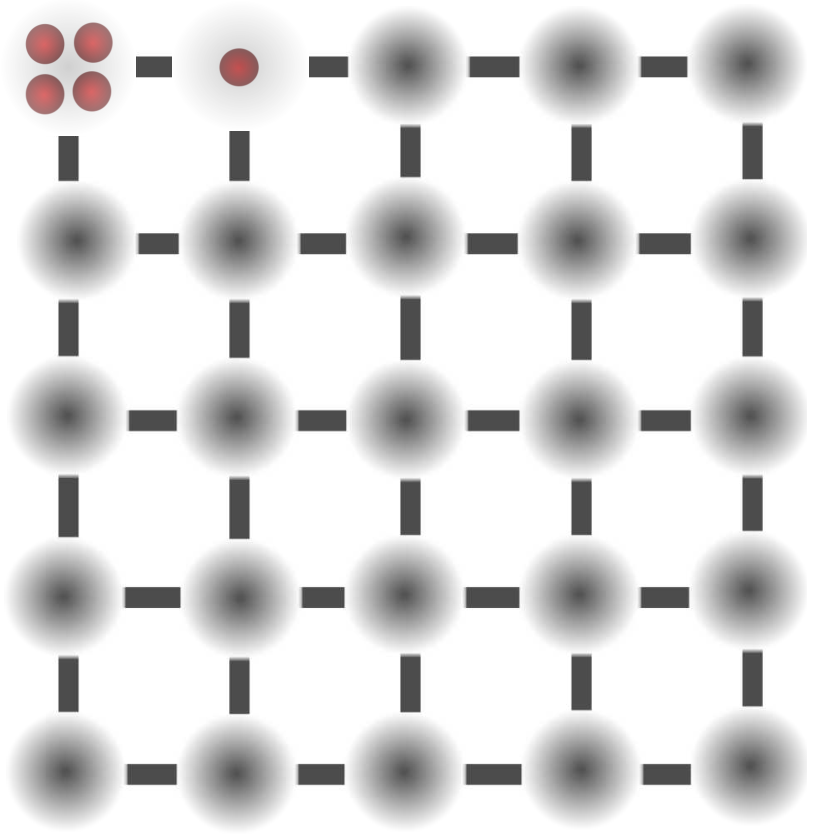}
        \label{fig:horiz}
    }
    \hspace*{-0.3cm}
    \subfigure[configuration $\ell_{\2}$]
    {
        \includegraphics[width=2.7cm]{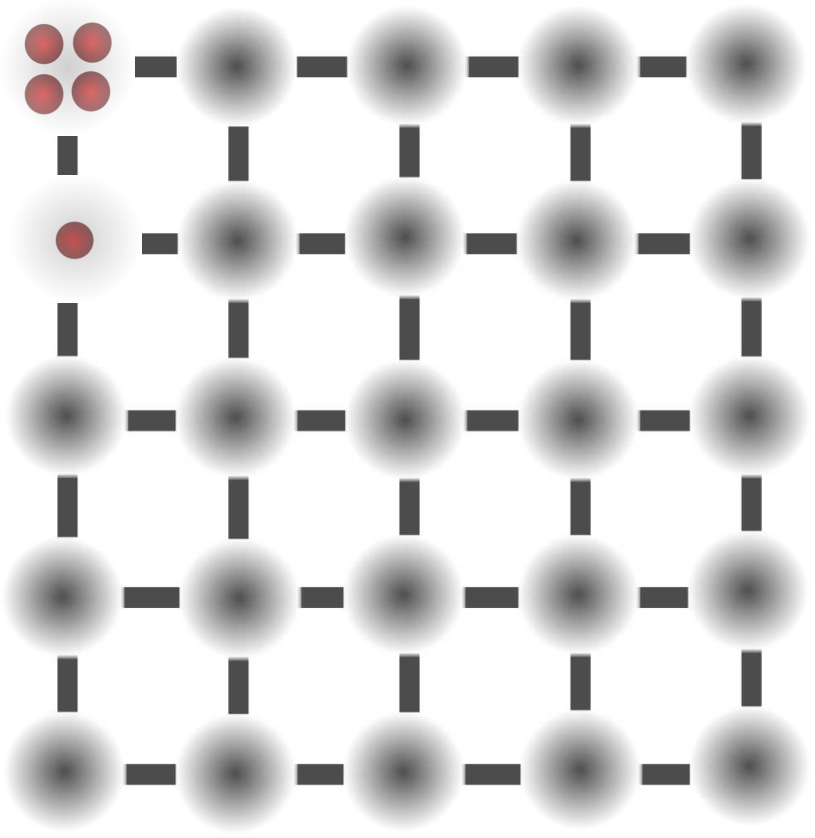}
        \label{fig:vert}
    }
    \caption{Initial GMP state and first step of five bosons quantum walks on lattice {I}.}
    \label{fig:step}
\end{figure}
The initial configuration together $\ell_{\0}$ with two configurations $\ell_{\1}$ and $\ell_{\2}$ created in the first step of the quantum walks are presented in Figs.~\ref{fig:step}. Subsequently at step 2 we have 
\begin{eqnarray}
\label{9.3}
  |\Psi_{\2}\rangle&=&\widehat{\mathcal{S}}^{2}|\Psi_{\0}\rangle\nonumber\\
  &=&\sum_{m}\frac{C_{\2\ell_{\0}}^{\0}e^{i\mbox{\tiny{$\frac{\pi}{2}$}}\tysl m-1\tysr}}{2^{\2}\mathcal{K}_{\0}}\sqrt{N(N-1)}|v_{m}\,{\bf n}_{\ell_{\4}}\rangle\nonumber\\
   &&\quad+\sum_{m}\frac{C_{\2\ell_{\0}}^{\0}e^{i\pi\tysl m-1\tysr}}{2^{\2}\mathcal{K}_{\0}}\sqrt{N(N-1)}|v_{m}\,{\bf n}_{\ell_{\0}}\rangle\nonumber\\
   &&\quad+\sum_{m}\frac{C_{\2\ell_{\0}}^{\0}+C_{\3\ell_{\0}}^{\0}e^{i\mbox{\tiny{$\frac{\3\pi}{2}$}}\tysl m-1\tysr}}{2^{\2}\mathcal{K}_{\0}}{N}|v_{m}\,{\bf n}_{\ell_{\6}}\rangle\nonumber\\
   &&\quad+\sum_{m}\frac{C_{\2\ell_{\0}}^{\0}e^{i\mbox{\tiny{$\frac{\pi}{2}$}}\tysl m-1\tysr}\delta_{k\2}}{2\mathcal{K}_{\0}}\sqrt{N}|v_{m}\,{\bf n}_{\ell_{\7}}\rangle\nonumber\\
   &&\quad+\sum_{m}\frac{(C_{\2\ell_{\0}}^{\0}+C_{\3\ell_{\0}}^{\0})e^{i\pi\tysl m-1\tysr}}{2^{\2}\mathcal{K}_{\0}}\sqrt{N}|v_{m}\,{\bf n}_{\ell_{\8}}\rangle\nonumber\\
   &&\quad+\sum_{m}\frac{C_{\3\ell_{\0}}^{\0}e^{\2i\pi\tysl m-1\tysr}}{2\mathcal{K}_{\0}}\sqrt{N}|v_{m}\,{\bf n}_{\ell_{\9}}\rangle,
\end{eqnarray}
where
\begin{widetext}
\begin{eqnarray}
&&|{\bf n}_{\ell_{\3}}\rangle=\sbox0{$\begin{array}{ccccc}
 N-2 & 2 & 0 & 0 & 0\\
 0 & 0 & 0 & 0 & 0\\
 0 & 0 & 0 & 0 & 0\\
 0 & 0 & 0 & 0 & 0\\
 0 & 0 & 0 & 0 & 0\end{array}$}
\mathopen{\resizebox{1.2\width}{\ht0}{$\Bigg|$}}
\usebox{0}
\mathclose{\resizebox{1.2\width}{\ht0}{$\Bigg\rangle$}},\quad|{\bf n}_{\ell_{\4}}\rangle=\sbox0{$\begin{array}{ccccc}
 N-2 & 0 & 0 & 0 & 0\\
 2 & 0 & 0 & 0 & 0\\
 0 & 0 & 0 & 0 & 0\\
 0 & 0 & 0 & 0 & 0\\
 0 & 0 & 0 & 0 & 0\end{array}$}
\mathopen{\resizebox{1.2\width}{\ht0}{$\Bigg|$}}
\usebox{0}
\mathclose{\resizebox{1.2\width}{\ht0}{$\Bigg\rangle$}},\quad
|{\bf n}_{\ell_{\0}}\rangle=\sbox0{$\begin{array}{ccccc}
 N & 0 & 0 & 0 & 0\\
 0 & 0 & 0 & 0 & 0\\
 0 & 0 & 0 & 0 & 0\\
 0 & 0 & 0 & 0 & 0\\
 0 & 0 & 0 & 0 & 0\end{array}$}
\mathopen{\resizebox{1.2\width}{\ht0}{$\Bigg|$}}
\usebox{0}
\mathclose{\resizebox{1.2\width}{\ht0}{$\Bigg\rangle$}},\nonumber\\
&&|{\bf n}_{\ell_{\6}}\rangle=\sbox0{$\begin{array}{ccccc}
 N-1 & 0 & 0 & 0 & 0\\
 0 & 1 & 0 & 0 & 0\\
 0 & 0 & 0 & 0 & 0\\
 0 & 0 & 0 & 0 & 0\\
 0 & 0 & 0 & 0 & 0\end{array}$}
\mathopen{\resizebox{1.2\width}{\ht0}{$\Bigg|$}}
\usebox{0}
\mathclose{\resizebox{1.2\width}{\ht0}{$\Bigg\rangle$}},\quad|{\bf n}_{\ell_{\7}}\rangle=\sbox0{$\begin{array}{ccccc}
 N-1 & 0 & 1 & 0 & 0\\
 0 & 0 & 0 & 0 & 0\\
 0 & 0 & 0 & 0 & 0\\
 0 & 0 & 0 & 0 & 0\\
 0 & 0 & 0 & 0 & 0\end{array}$}
\mathopen{\resizebox{1.2\width}{\ht0}{$\Bigg|$}}
\usebox{0}
\mathclose{\resizebox{1.2\width}{\ht0}{$\Bigg\rangle$}},\quad|{\bf n}_{\ell_{\8}}\rangle=\sbox0{$\begin{array}{ccccc}
 N-1 & 0 & 0 & 0 & 0\\
 0 & 0 & 0 & 0 & 0\\
 1 & 0 & 0 & 0 & 0\\
 0 & 0 & 0 & 0 & 0\\
 0 & 0 & 0 & 0 & 0\end{array}$}
\mathopen{\resizebox{1.2\width}{\ht0}{$\Bigg|$}}
\usebox{0}
\mathclose{\resizebox{1.2\width}{\ht0}{$\Bigg\rangle$}}.\nonumber
\end{eqnarray}
\end{widetext}
 In our study of many-particle quantum walks on lattice we measure the time dependent vertex particle density $\langle{n}_{\alpha}\sml{r}\smr\rangle$ and the vertices counting statistics. We use the counting statistics as a detection tool of the presence of the initial configuration in the state of the system that gives us the ETH probing. We will only present the measurement of all $5$ particles on each vertex at the selected time steps. In fact,  simultaneous counting of five quantum walkers on vertex $\alpha=1$ coincides with the presence of initial configuration (\ref{4}) in the linear combination of configurations forming  the GMP state after $r$ steps.  In addition, we also measure the entropy of the configuration (\ref{4}) and its temperature. The results for these two observables will be presented in the next section. 

Let us discuss the results of simulations obtained after $r=200$ steps for $N=5$ bosons and $N=5$ fermions quantum walks on lattice {I} and lattice {II}. We evaluate vertex particle distribution according to the formula 
\begin{equation}
 \label{10}
\langle{n}_{\alpha}\sml{r}\smr\rangle=\langle\Psi_{r}|\widehat{\mathbf{\Uppsi}}^{\dagger}\sml{x}_{\alpha}\smr\widehat{\mathbf{\Uppsi}}\sml{x}_{\alpha}\smr|\Psi_{r}\rangle
\end{equation}
and the results of these simulations are presented in Figs \ref{fig:density}. The algorithm of the computation is presented in the Appendix. Plots in the left column show the bosons quantum walks on the lattices {I} and {II}, respectively, see Fig. ~\ref{fig:density1} and Fig. ~\ref{fig:density3}. In these figures we observe that the boson bunching effects on particle densities have maximum values around 0.8 on both lattices. For fermions the particle densities  plotted in the right column take values below 0.6. This shows the repulsion between fermions on the same vertex for lattices {I} and {II}, compare Fig. ~\ref{fig:density3} and Fig. ~\ref{fig:density4}.   
\vspace*{-0.1cm}
\begin{figure}[htb]
    \centering
       \subfigure[5 bosons on lattice {I}]
    {
        \includegraphics[width=4.2cm, height=3cm]{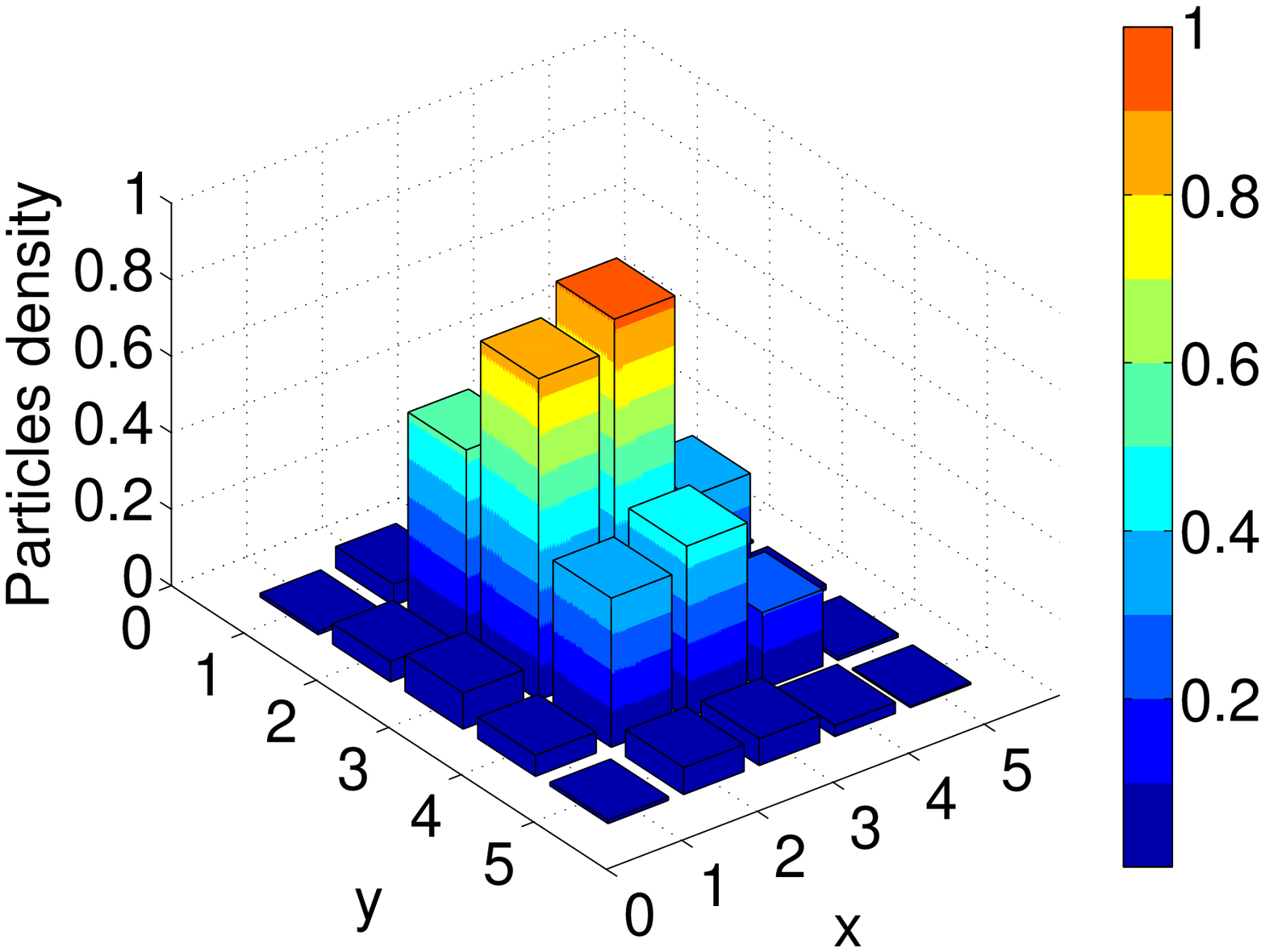}
         \label{fig:density1}
    }
    \hspace*{-0.5cm}
    \subfigure[5 fermions on lattice {I} ]
    {
        \includegraphics[width=4.2cm, height=3cm]{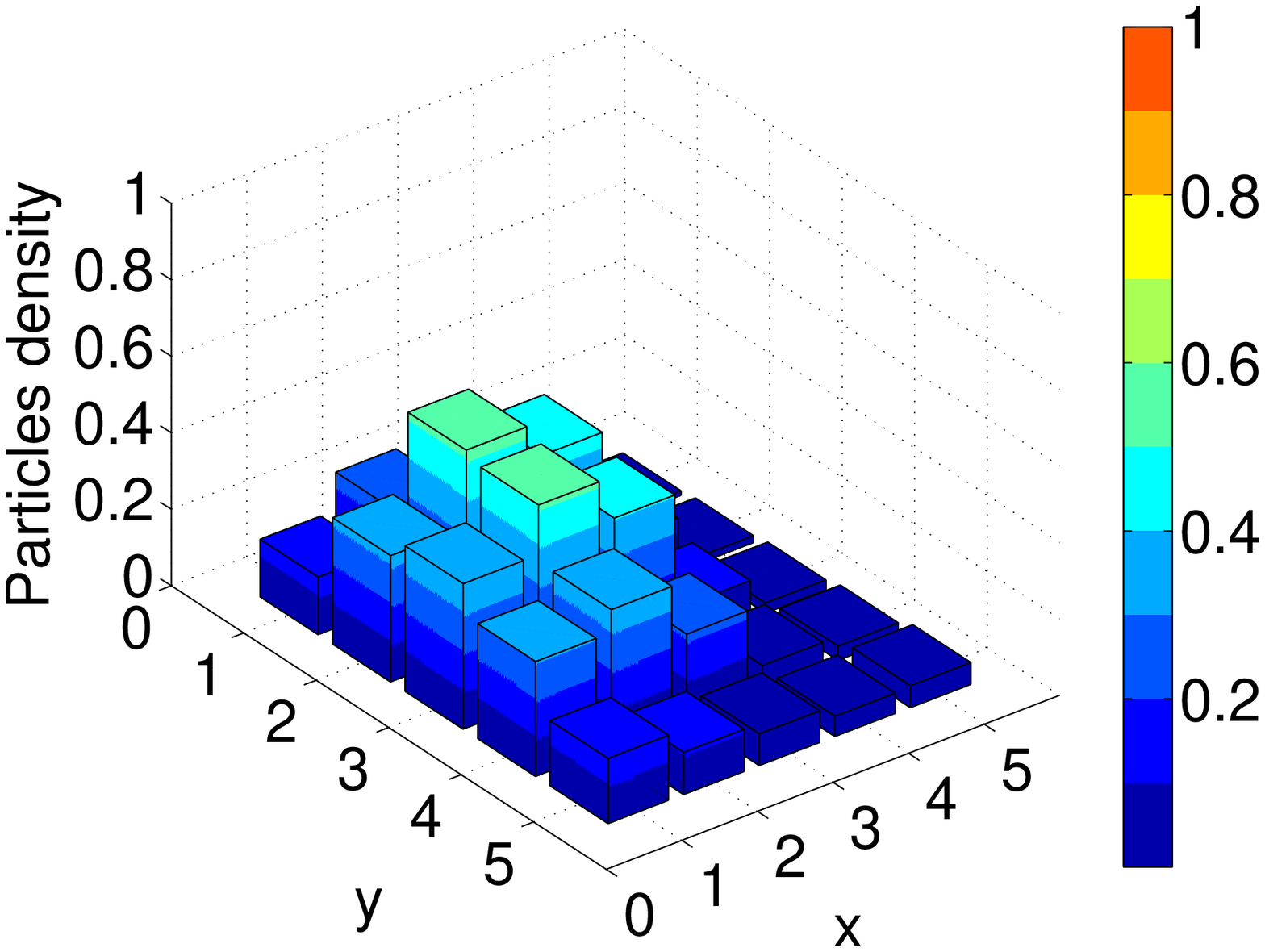}
         \label{fig:density2}
    }\\
    \subfigure[5 bosons on lattice {II}]
    {
        \includegraphics[width=4.2cm, height=3cm]{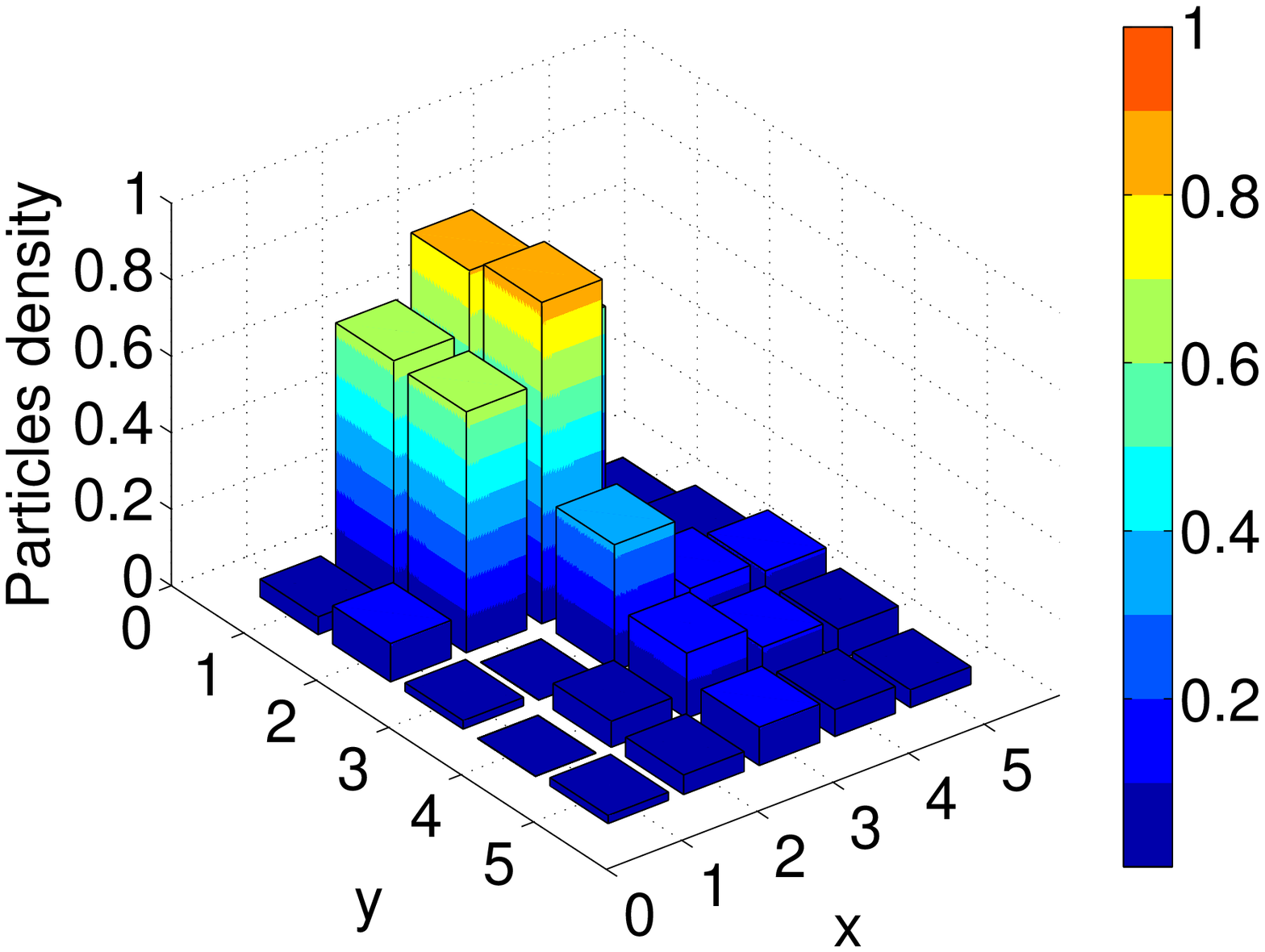}
         \label{fig:density3}
    }
    \hspace*{-0.5cm}
     \subfigure[5 fermions on lattice {II} ]
    {
        \includegraphics[width=4.2cm, height=3cm]{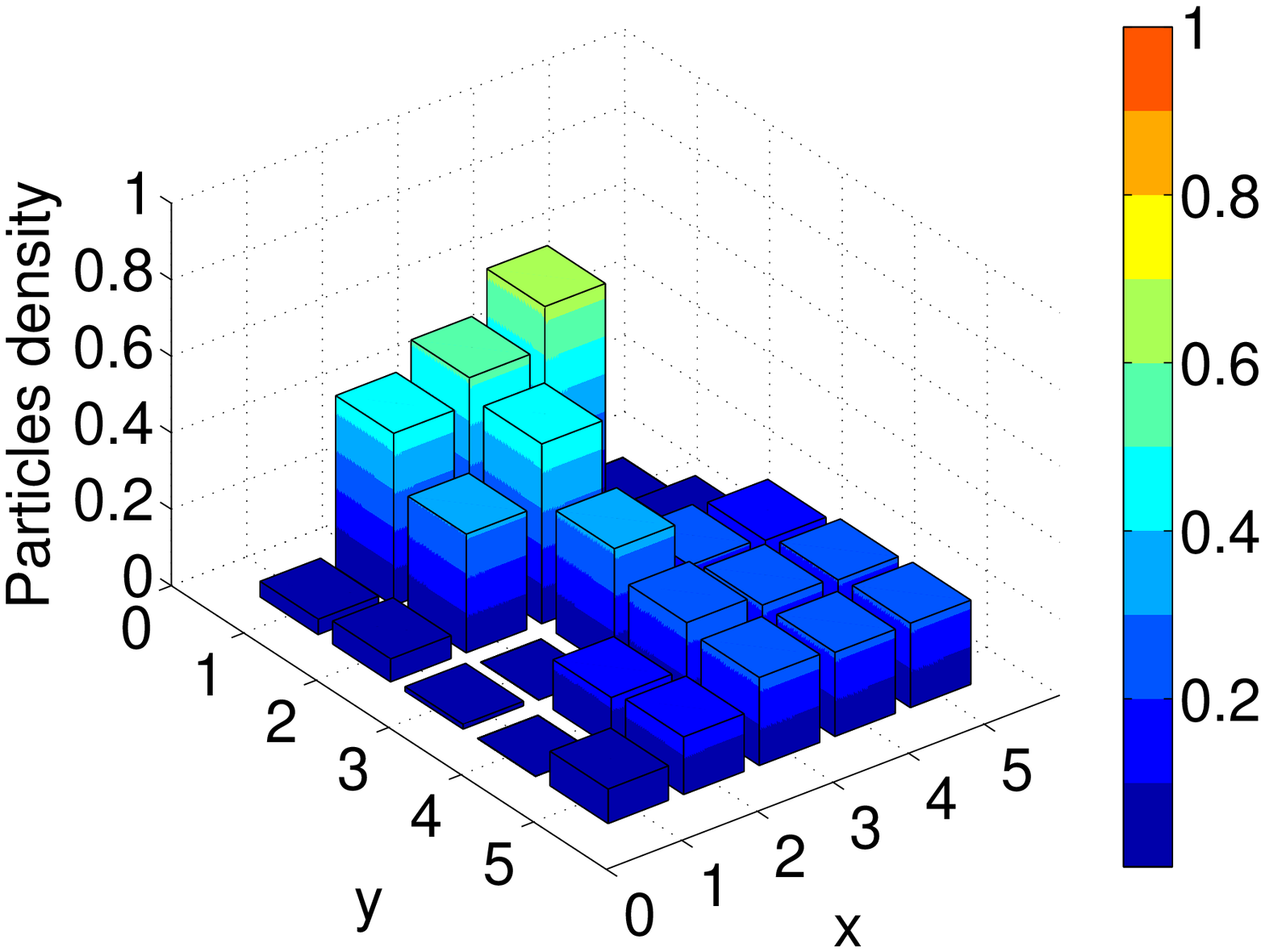}
         \label{fig:density4}
    }
    \caption{Vertices particle density $\langle{n}_{\alpha}\sml{r}\smr\rangle$ for $r=200$ steps of five bosons and five fermions  quantum walks on lattices {I} and {II}.}
    \label{fig:density}
\end{figure}
It is worth mentioning that two vertices with the same number of fermions are treated as identical because all the populations occupy the lowest available modes. Therefore during the shifting the highest mode fermion is shifted and if the destination vertex has exactly the same number of fermions as the vertex with the incoming fermion, these two fermions are in exactly the same mode and therefore such a shifting is forbidden. 
\begin{figure*}
\begin{center}
      \subfigure[Bosons $r=10$]
      {
      \includegraphics[width=4cm, height=3cm]{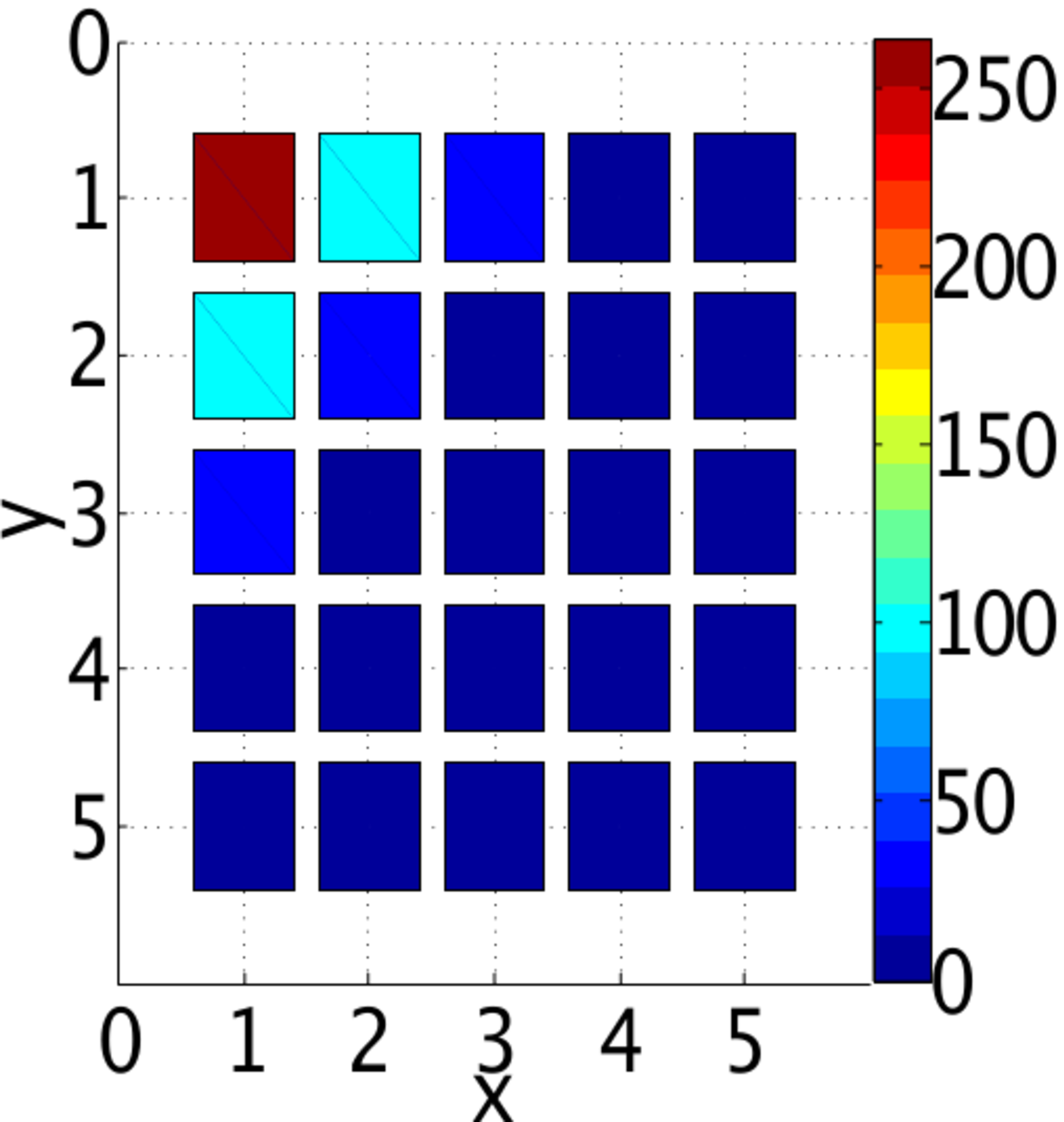}
      \label{fig:gridgr11}
      }
      \hspace*{-0.2cm}
       \subfigure[Bosons $r=10$]
     {
      \includegraphics[width=4cm, height=3cm]{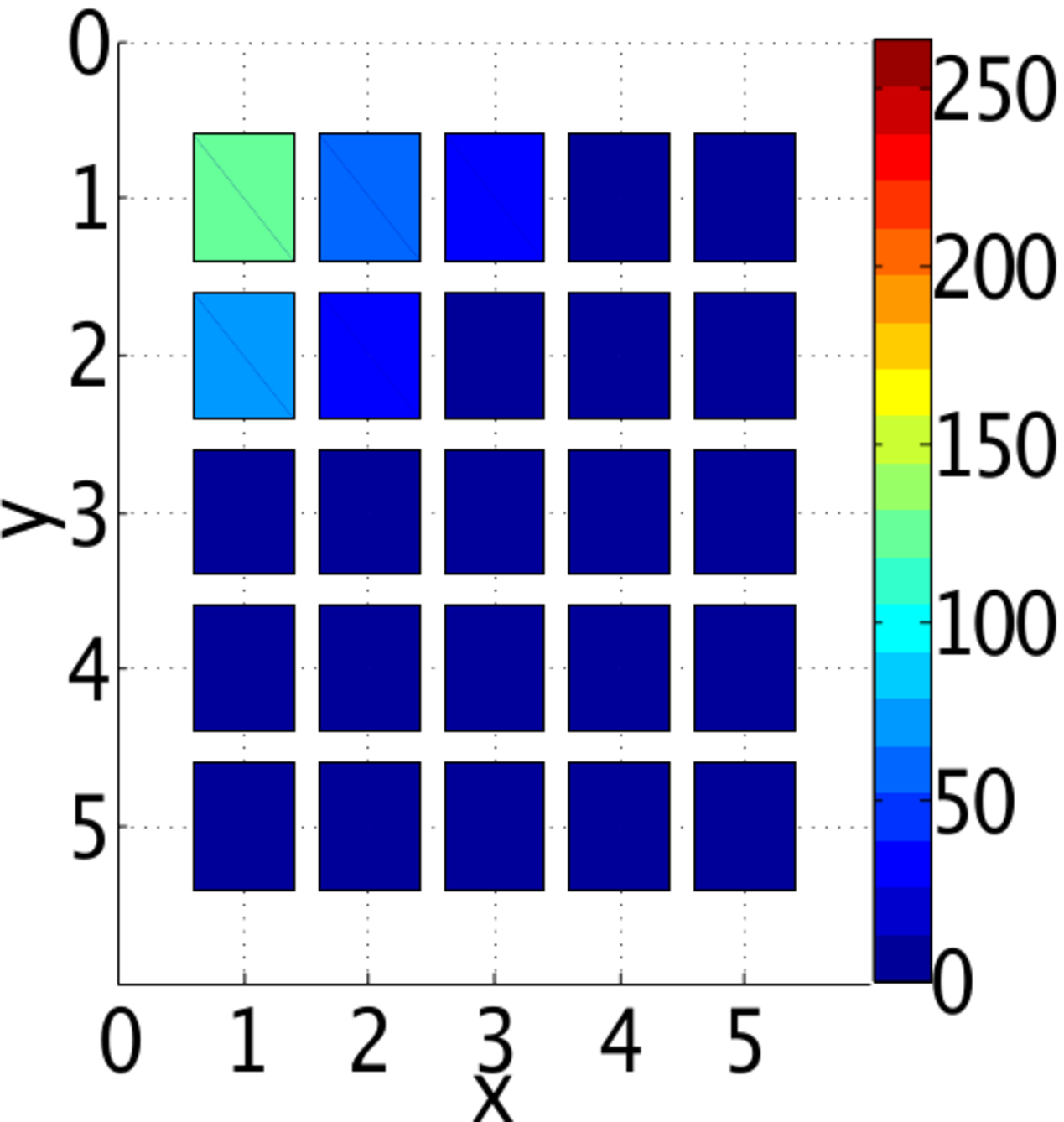}
      \label{fig:gridgr12}
     }
     \hspace*{-0.2cm}
      \subfigure[Fermions $r=10$]
      {
      \includegraphics[width=4cm, height=3cm]{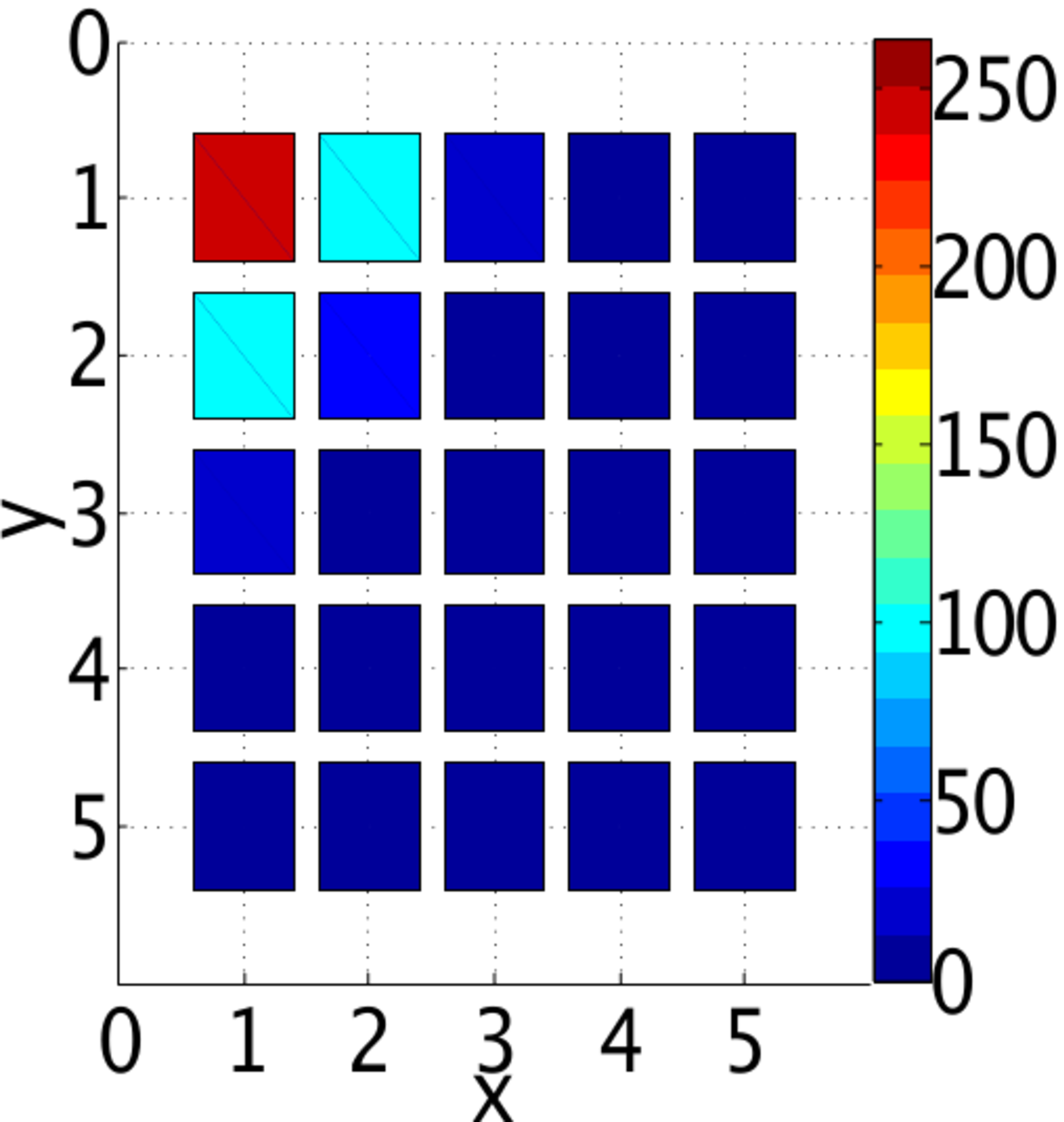}
      \label{fig:gridgr13}
      }
      \hspace*{-0.2cm}
       \subfigure[Fermions $r=10$]
     {
      \includegraphics[width=4cm, height=3cm]{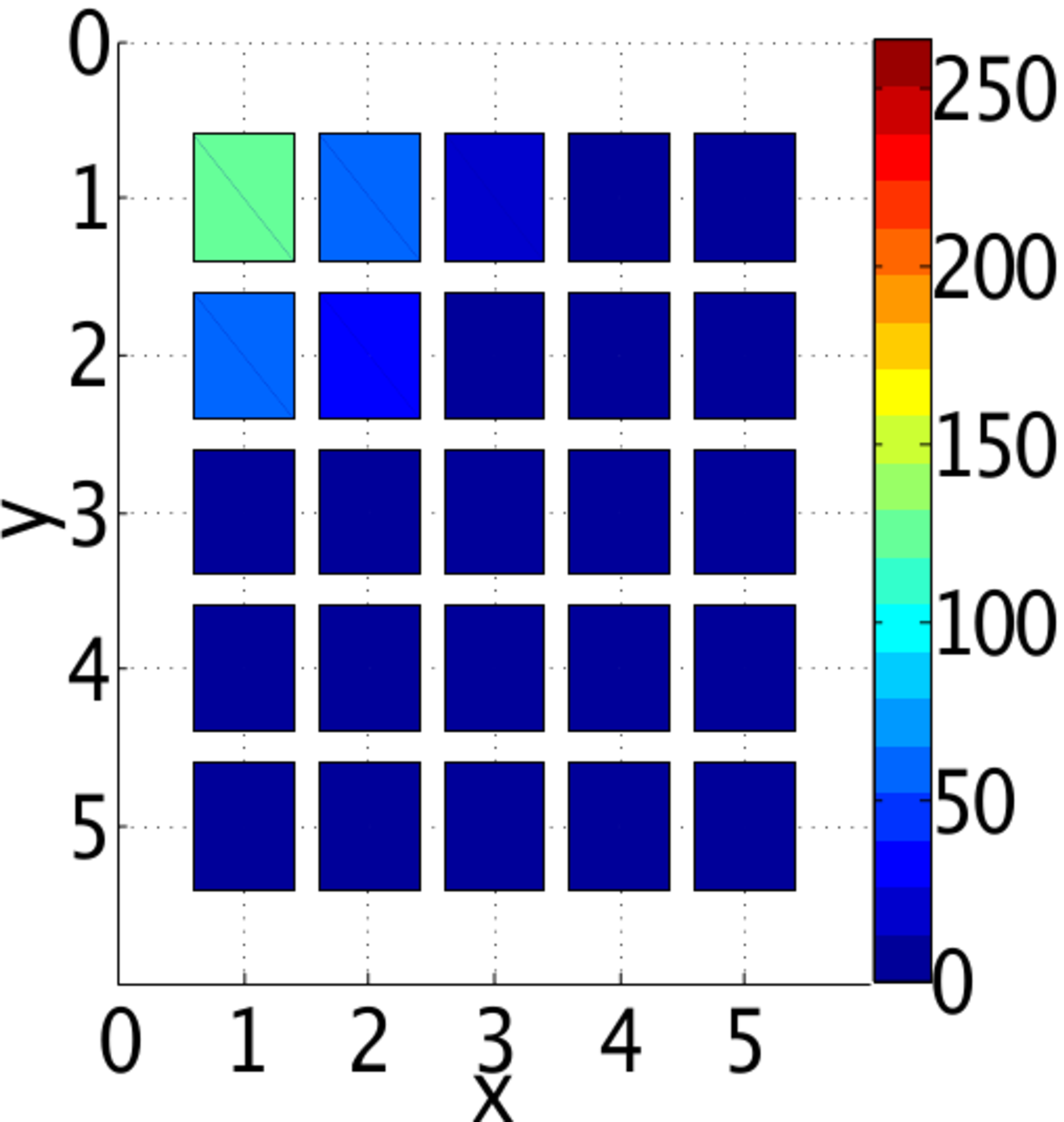}
      \label{fig:gridgr14}
     }
    \\
    \vspace*{-0.1cm}
     \subfigure[Bosons $r=50$]
      {
      \includegraphics[width=4cm, height=3cm]{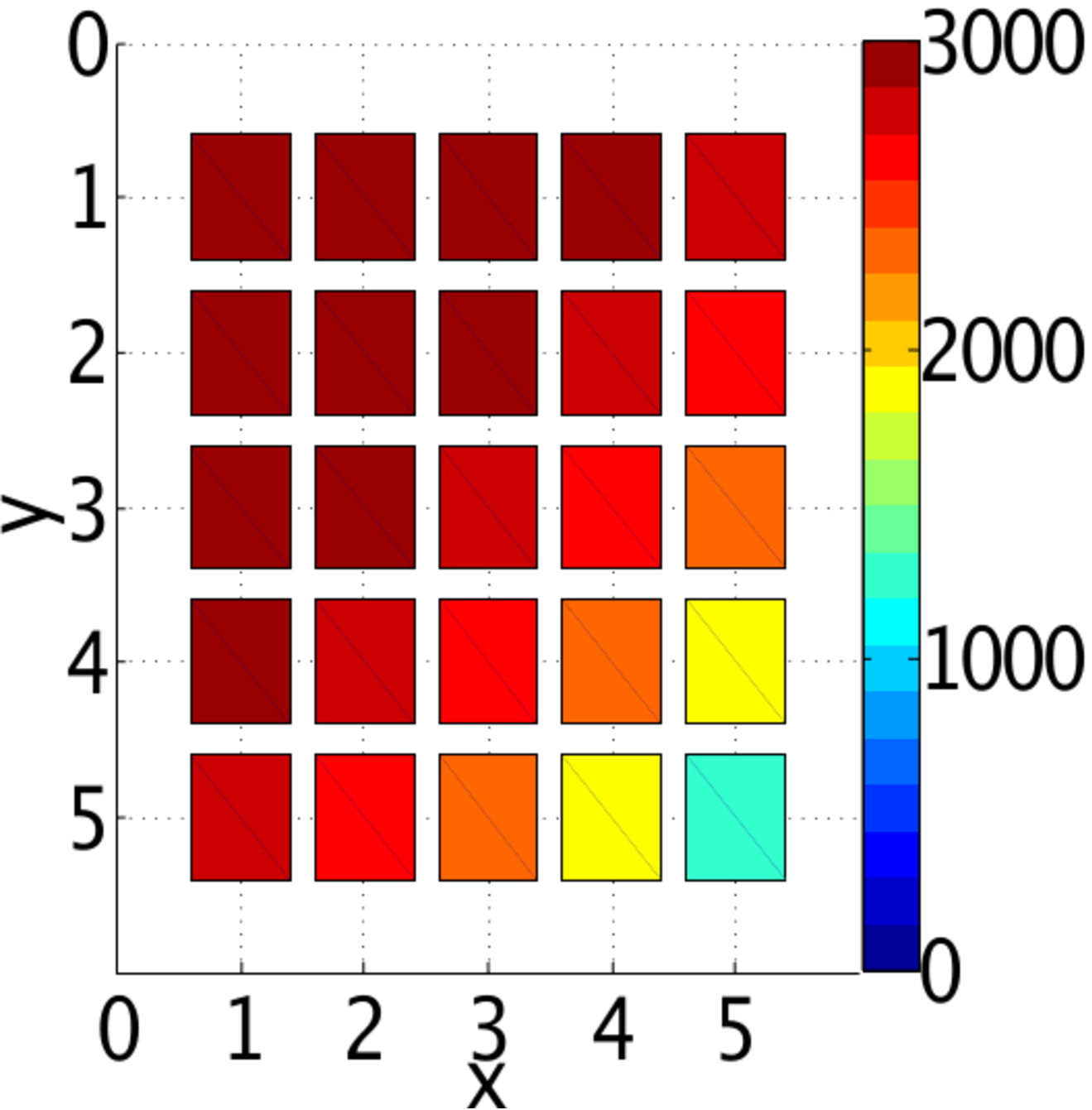}
      \label{fig:gridgr21}
      }
      \hspace*{-0.2cm}
       \subfigure[Bosons $r=50$]
     {
      \includegraphics[width=4cm, height=3cm]{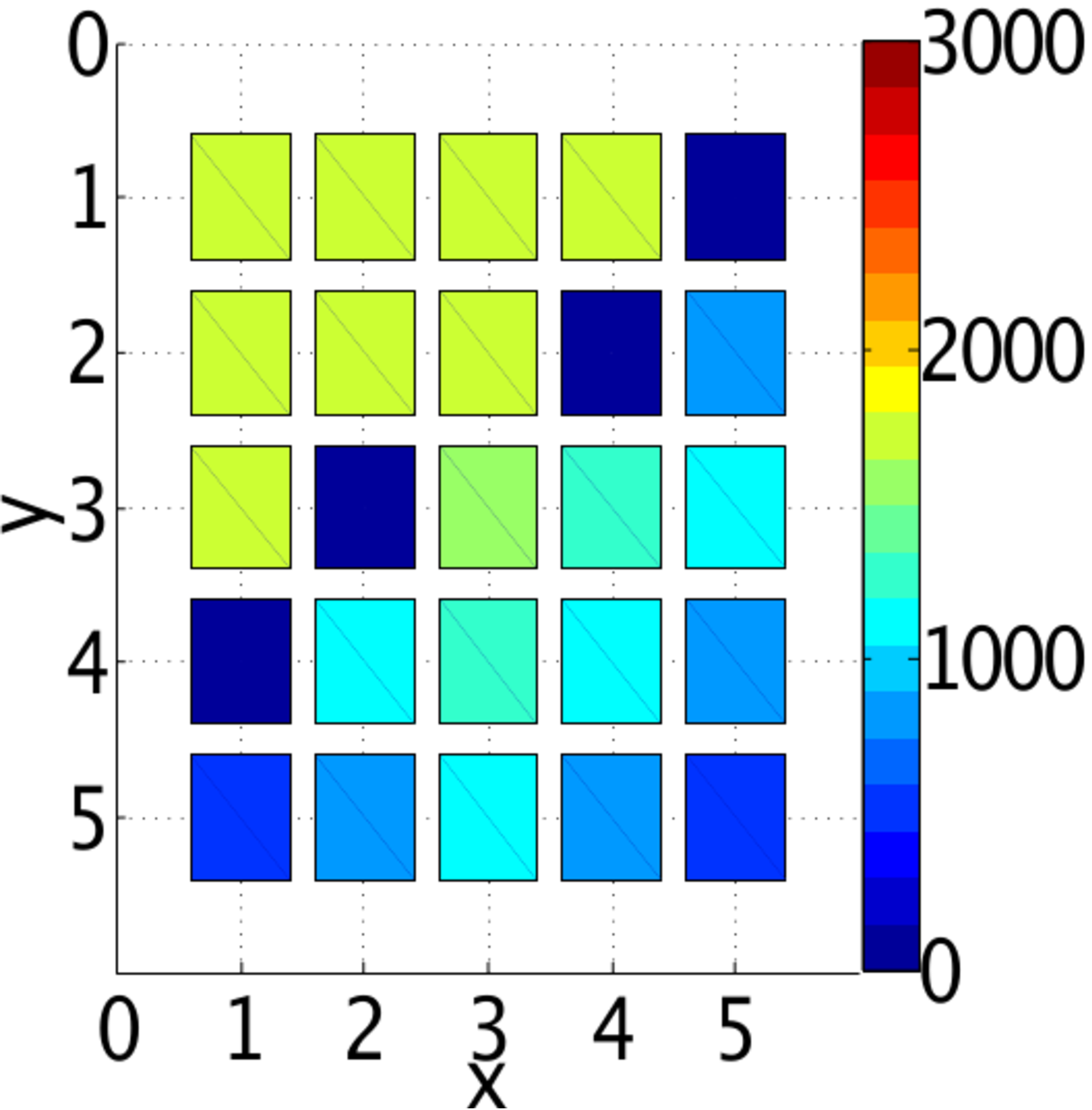}
      \label{fig:gridgr22}
     }
     \hspace*{-0.2cm}
      \subfigure[Fermions $r=50$]
      {
      \includegraphics[width=4cm, height=3cm]{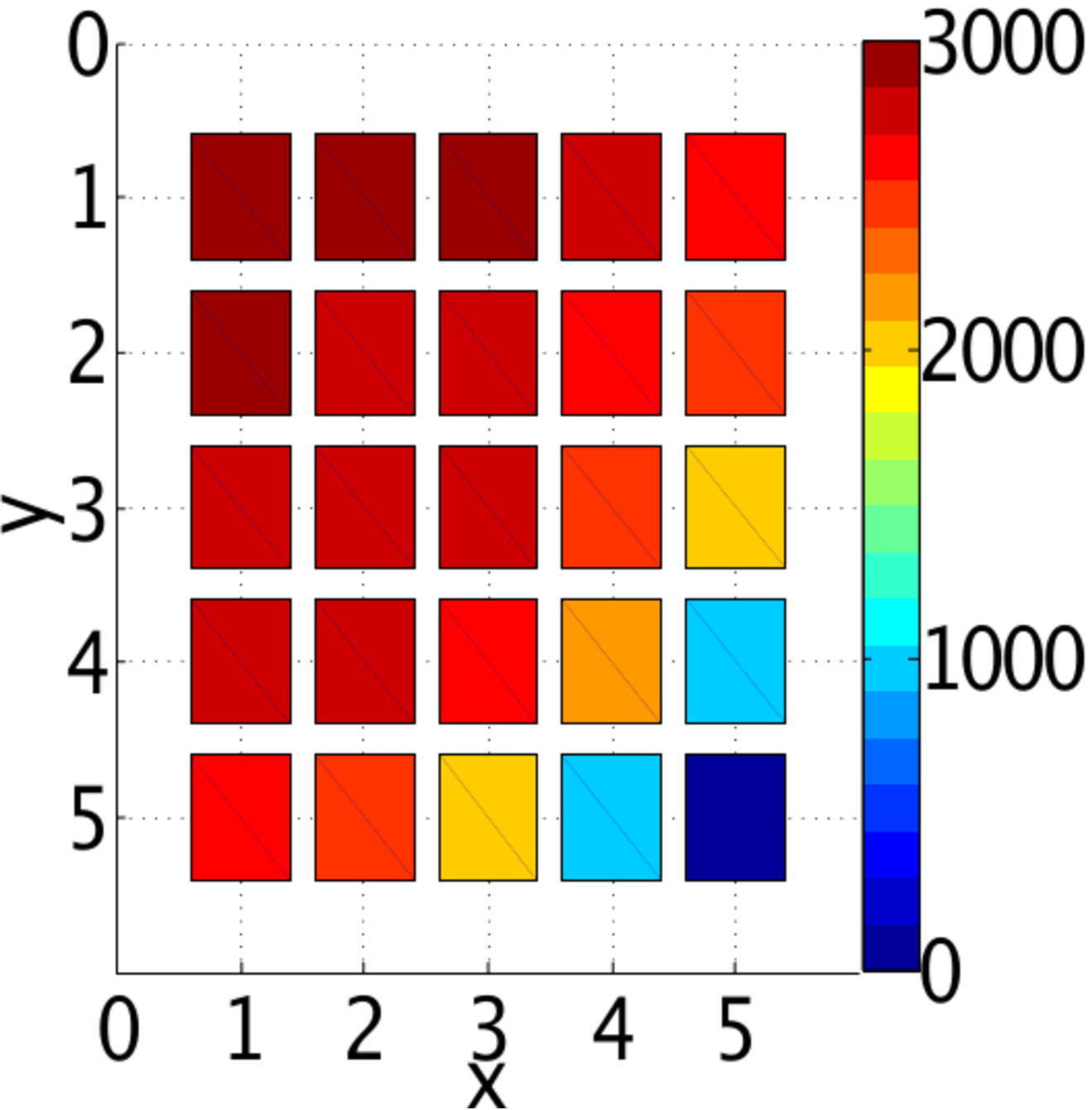}
      \label{fig:gridgr23}
      }
      \hspace*{-0.2cm}
       \subfigure[Fermions $r=50$]
     {
      \includegraphics[width=4cm, height=3cm]{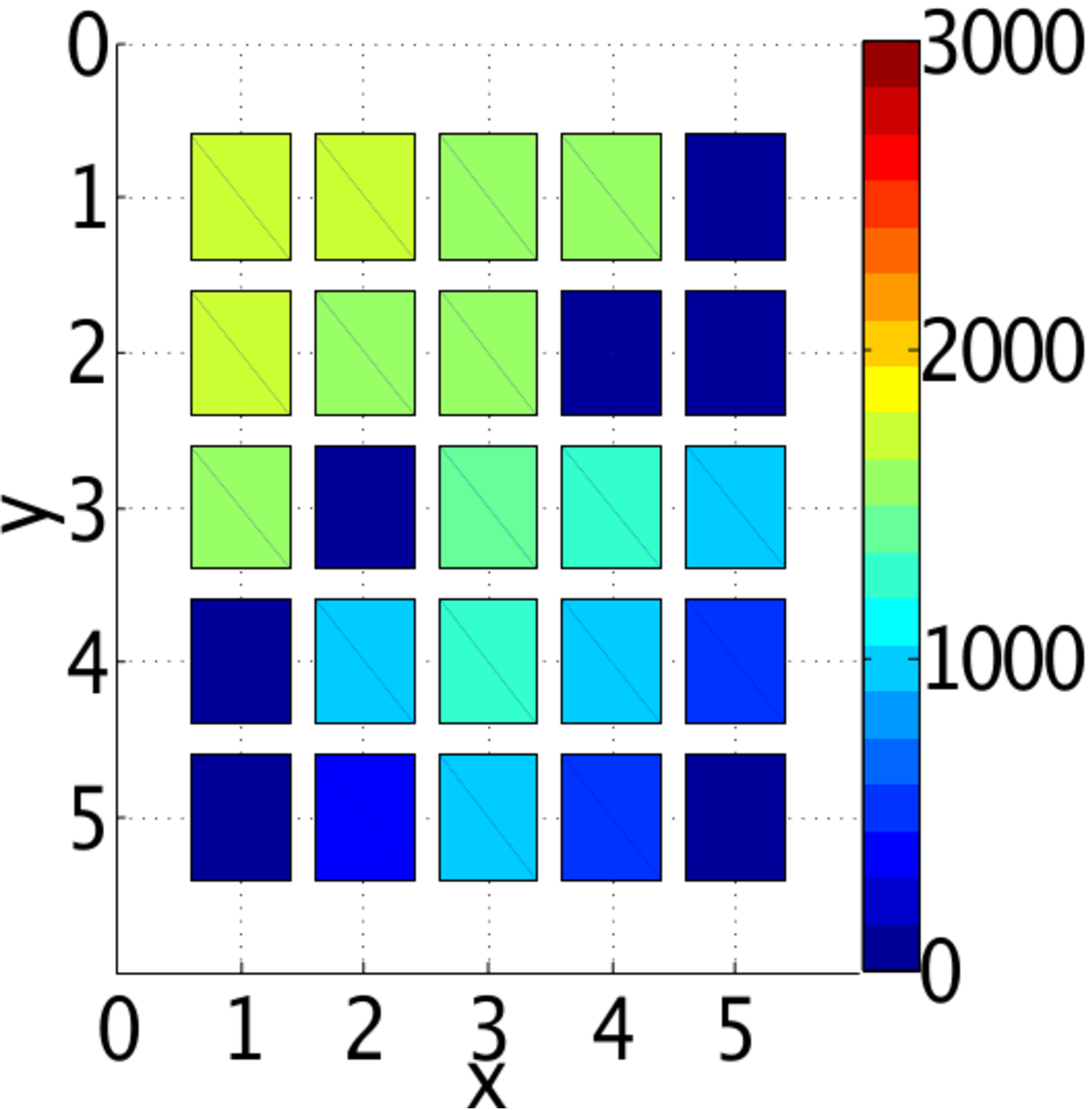}
      \label{fig:gridgr24}
     }
    \\
     \vspace*{-0.1cm}
      \subfigure[Bosons $r=100$]
      {
      \includegraphics[width=4cm, height=3cm]{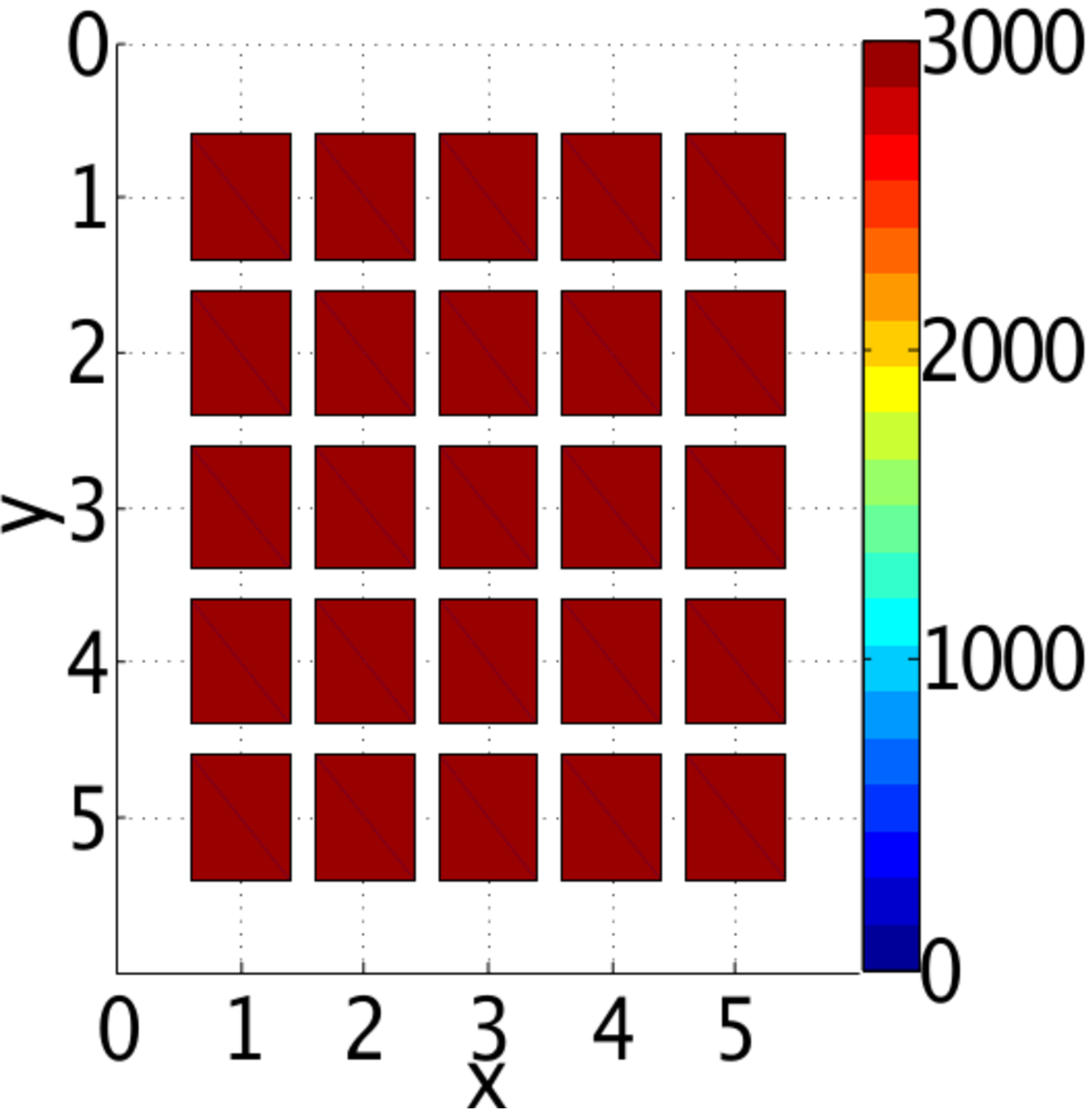}
      \label{fig:gridgr31}
      }
      \hspace*{-0.2cm}
       \subfigure[Bosons $r=100$]
     {
      \includegraphics[width=4cm, height=3cm]{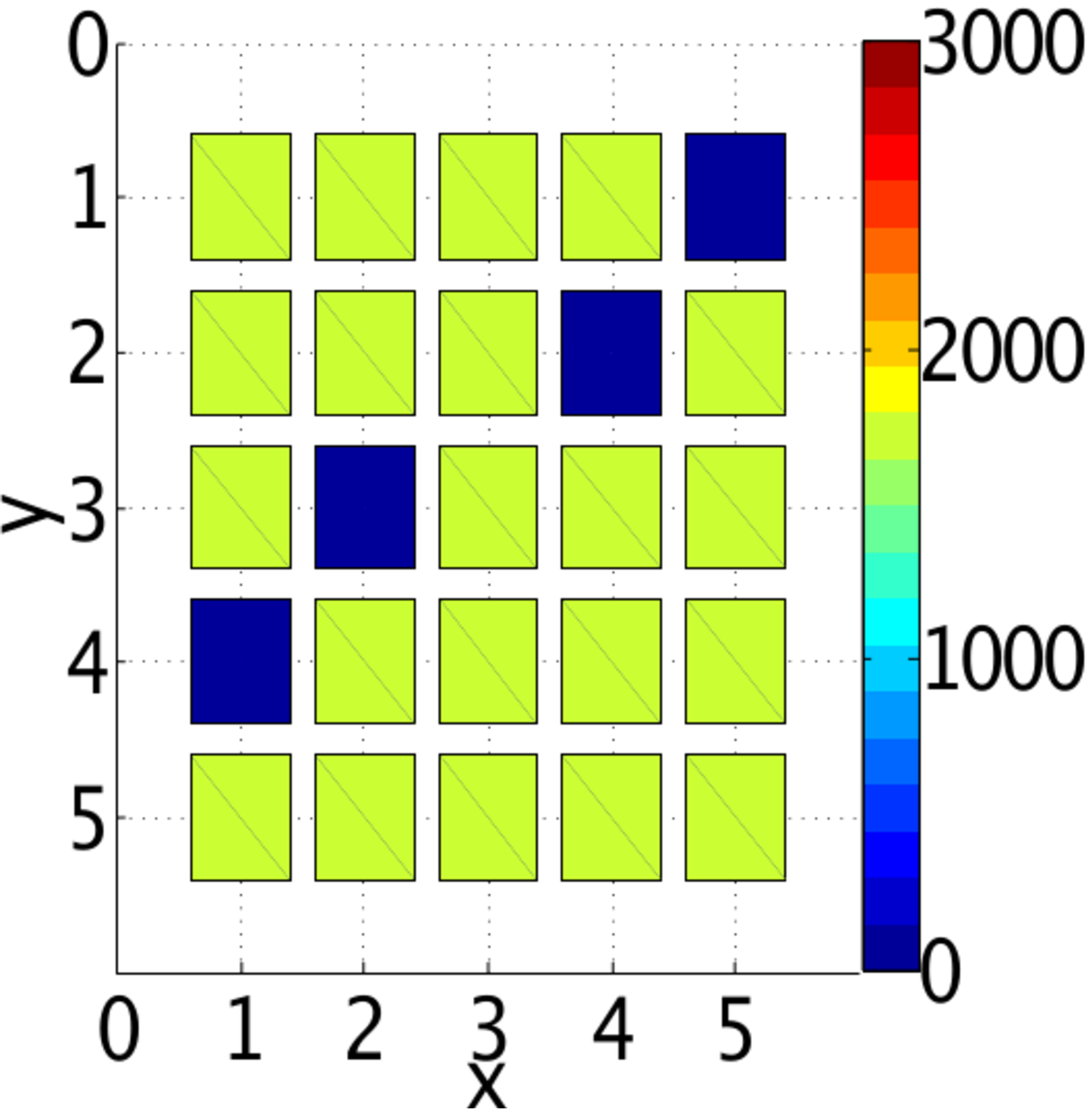}
      \label{fig:gridgr32}
     }
     \hspace*{-0.2cm}
      \subfigure[Fermions $r=100$]
      {
      \includegraphics[width=4cm, height=3cm]{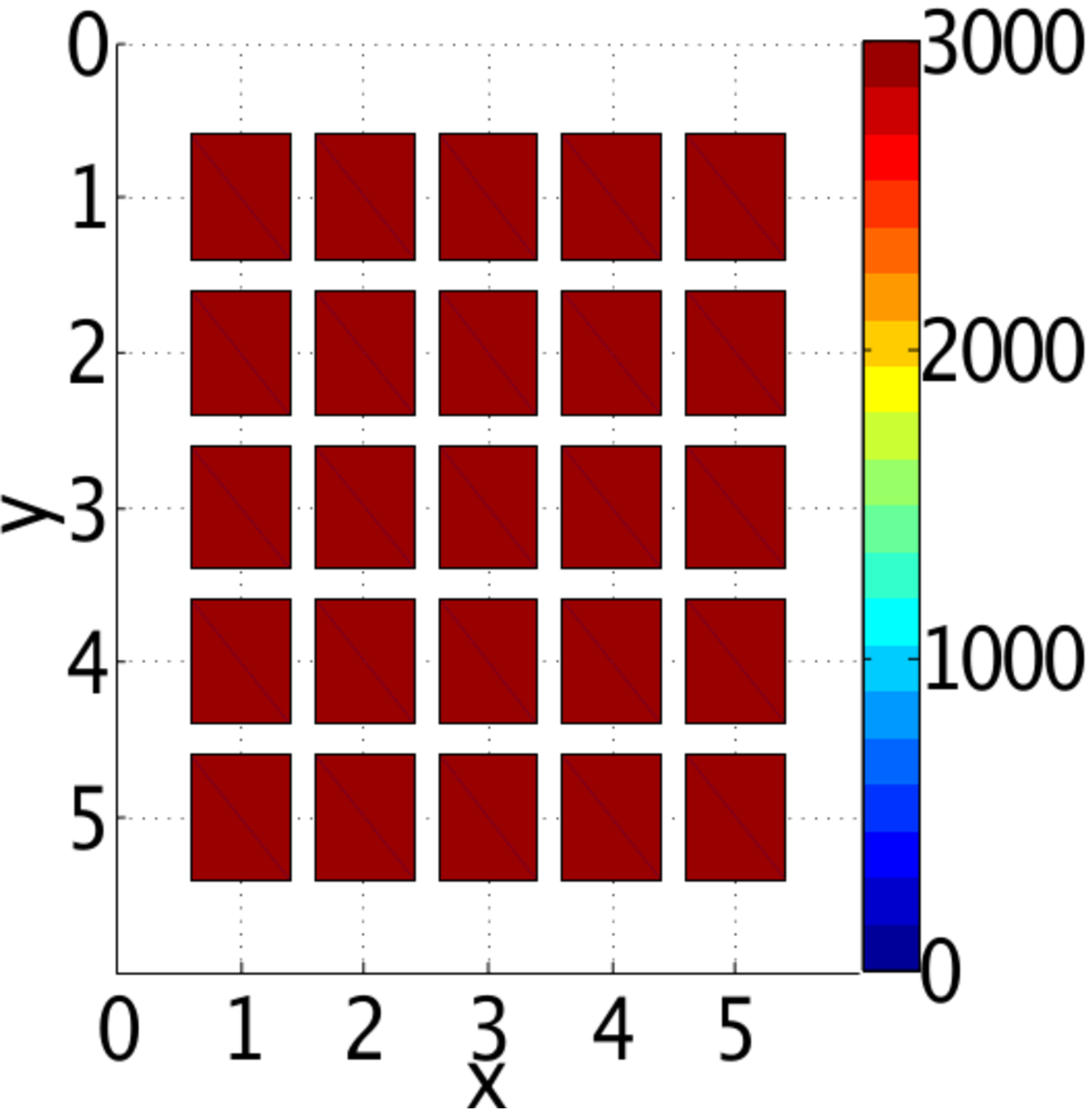}
      \label{fig:gridgr33}
      }
      \hspace*{-0.2cm}
       \subfigure[Fermions $r=100$]
     {
      \includegraphics[width=4cm, height=3cm]{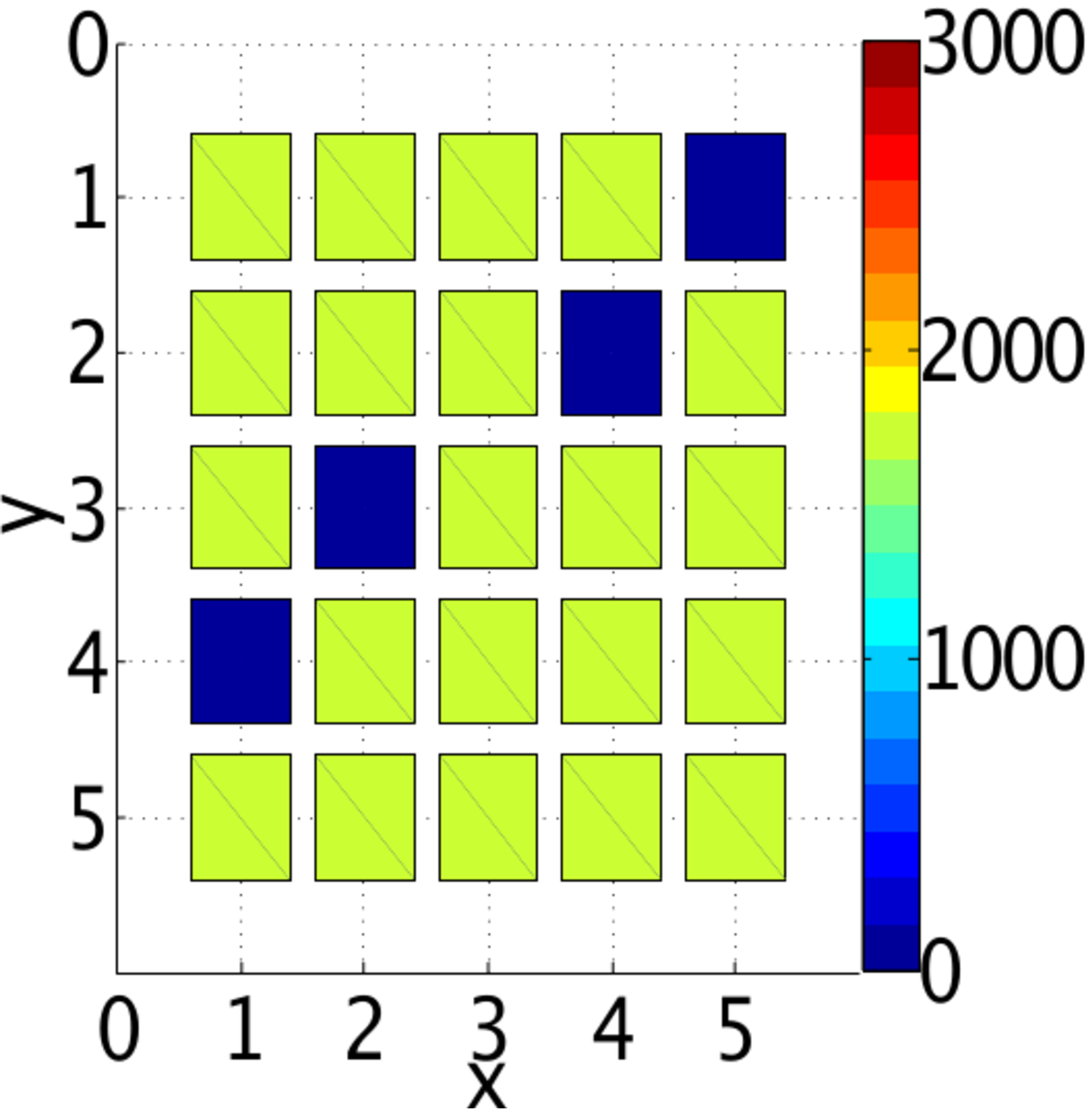}
      \label{fig:gridgr34} 
      }    
      \\
      \vspace*{-0.1cm}
      \subfigure[Bosons $r=200$]
      {
      \includegraphics[width=4cm, height=3cm]{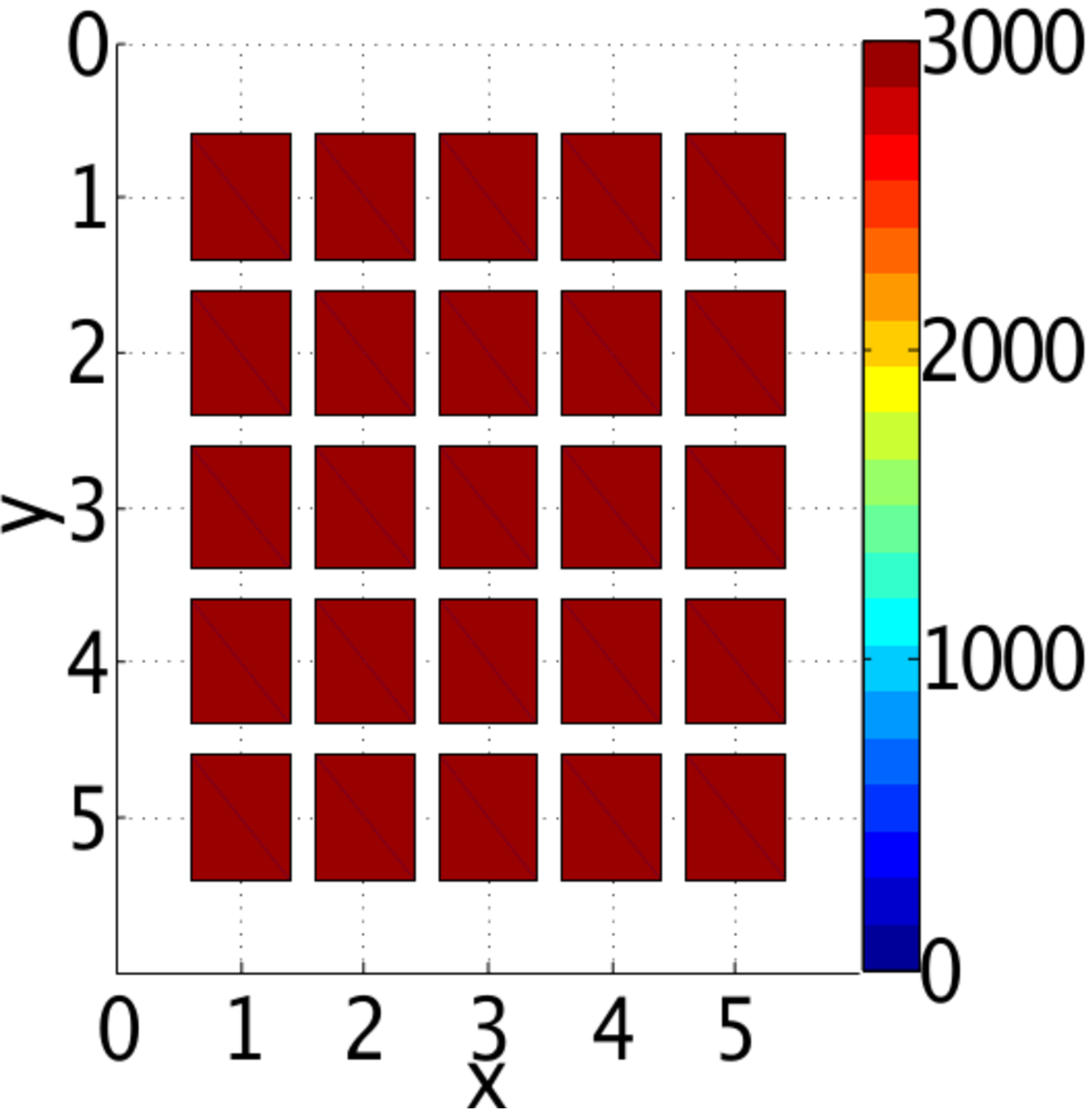}
      \label{fig:gridgr41}
      }
      \hspace*{-0.2cm}
       \subfigure[Bosons $r=200$]
     {
      \includegraphics[width=4cm, height=3cm]{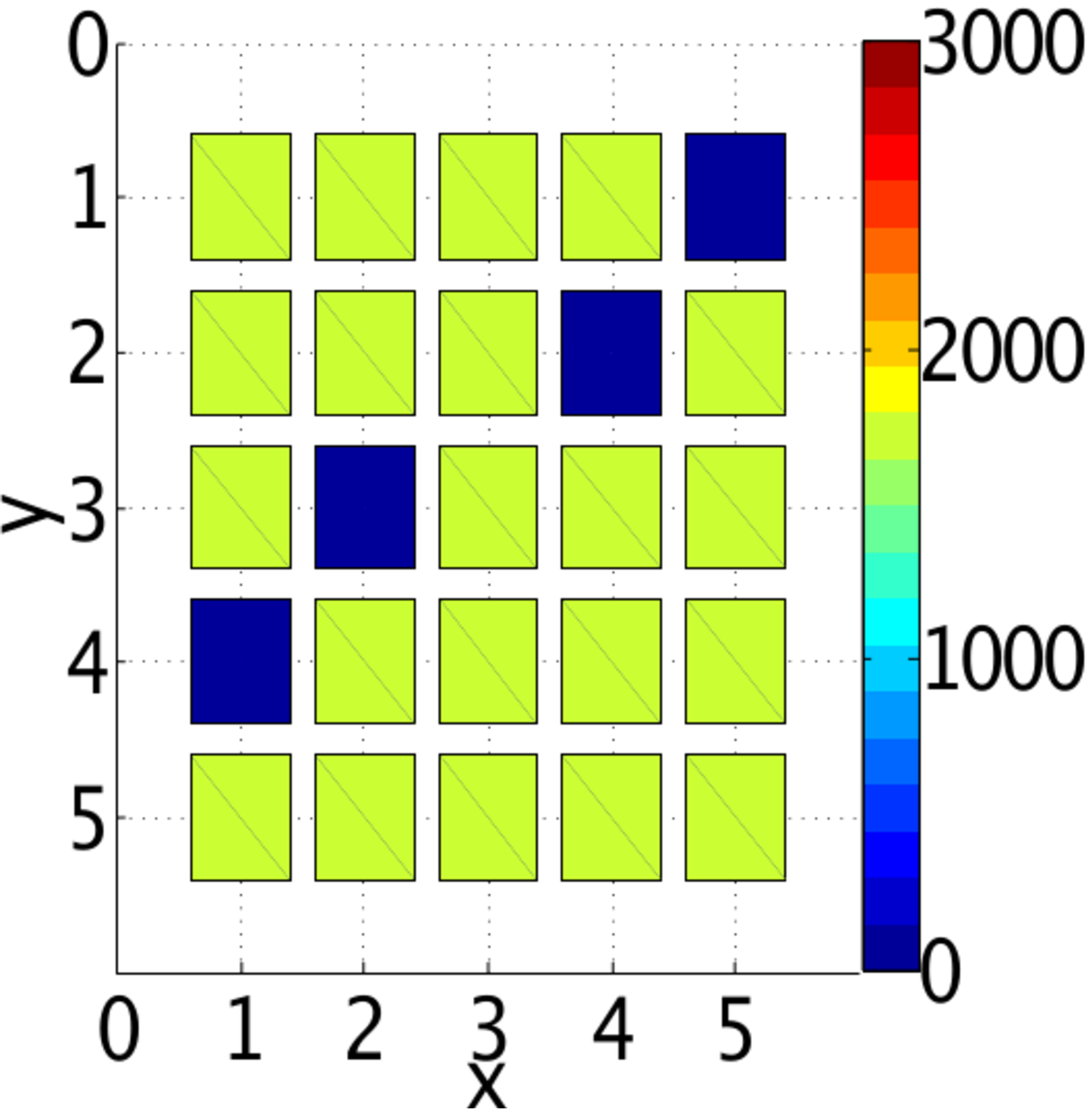}
      \label{fig:gridgr42}
     }
     \hspace*{-0.2cm}
      \subfigure[Fermions $r=200$]
      {
      \includegraphics[width=4cm, height=3cm]{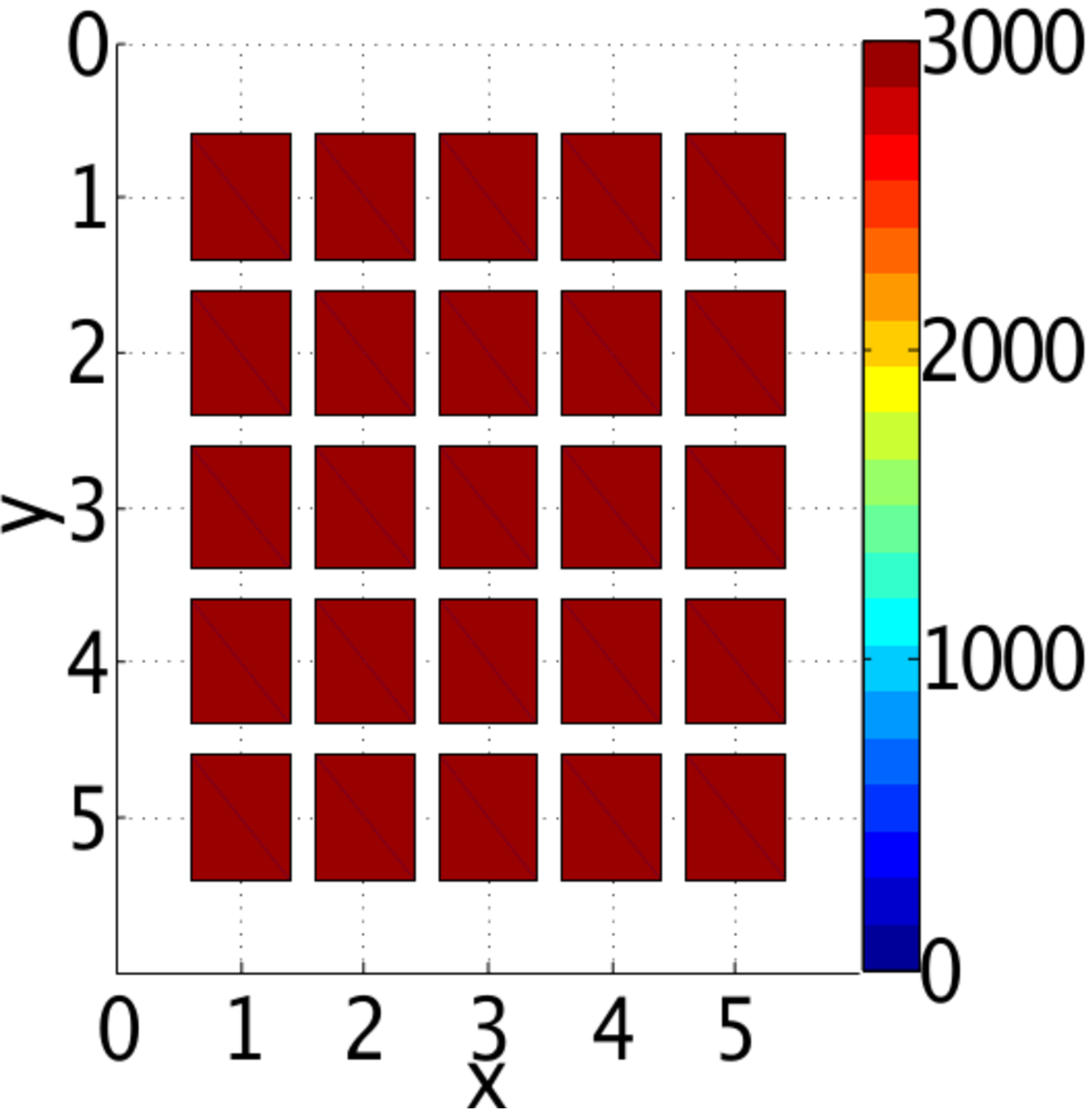}
      \label{fig:gridgr43}
      }
      \hspace*{-0.2cm}
       \subfigure[Fermions $r=200$]
     {
      \includegraphics[width=4cm, height=3cm]{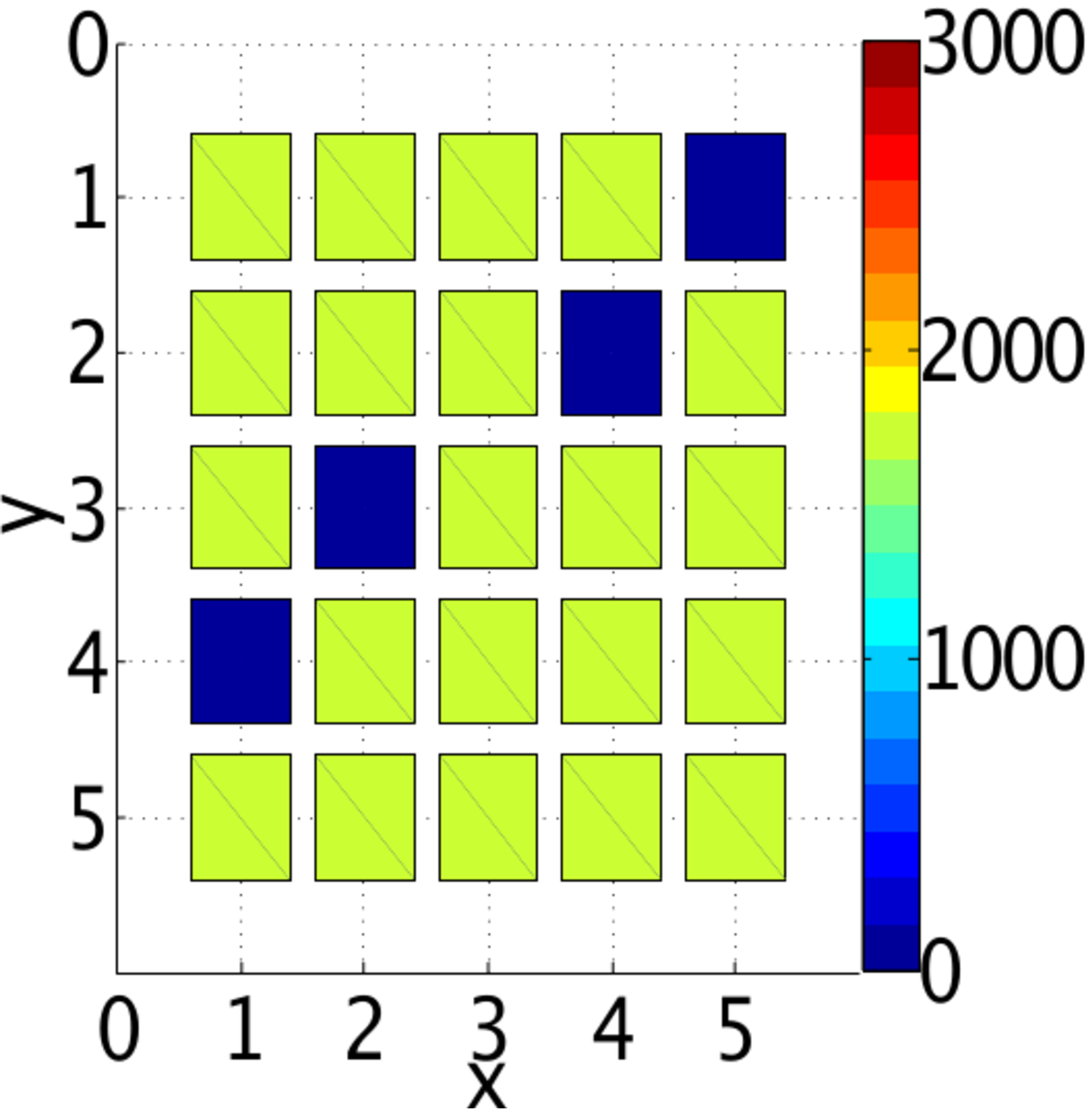}
      \label{fig:gridgr44} 
      }       
      \caption{Selected time steps $r$ simulations of the counting $P_{2}^{r}$  when two particles are detected on a vertex. Figs. ~\ref{fig:gridgr11}, \ref{fig:gridgr21}, \ref{fig:gridgr31}, \ref{fig:gridgr41} represent $N=5$ bosons on lattice {I}, Figs \ref{fig:gridgr12}, \ref{fig:gridgr22}, \ref{fig:gridgr32}, \ref{fig:gridgr42} represent $N=5$ bosons on lattice {II}, Figs. ~\ref{fig:gridgr13}, \ref{fig:gridgr23}, \ref{fig:gridgr33}, \ref{fig:gridgr43} concern $N=5$ fermions on lattice {I} and Figs. ~\ref{fig:gridgr14}, \ref{fig:gridgr24}, \ref{fig:gridgr34}, \ref{fig:gridgr44} concern $N=5$ fermions on lattice {II}.}
    \label{fig:gridgr}
\end{center}
\end{figure*}

To detect the quantum walks thermalization, we will use the vertices counting statistics that is an example of a local observable.  To illustrate this notion we consider a particle detector on vertex $\alpha$ fine tuned to record a count only when $n_{\alpha}$ walkers report on  that vertex. This counting brings together the combinatoric problem  of distributing $N$ indistinguishable bosons/fermions among $M^{\2}$ vertices and the probability that a given eigenvector of the configurations Hilbert space has a non-zero amplitude. Let us measure the number of times that $n_{\alpha}\leq N$ of particles are located on vertex $\alpha$.  A such event is defined by the quantity  $\frac{D\tyl \mbox{\tiny{$N$}}-n_{\alpha},\mbox{\tiny{$M^{\2}-1$}}\tyr}{M^{\2}D\tyl\mbox{\tiny{$N,M^{\2}$}}\tyr}$  obtained from the combinatoric problem of distributing of $N-n_{\alpha}$ indistinguishable particles in $M^{\2}-1$ vertices, where $D\sml\mbox{\small{$N,M^{\2}$}}\smr$ is defined in Eq.(\ref{1}). This event must coincide with the configuration $|{\bf n}_{\ell}\rangle$ having nonzero amplitude  \cite{Pierrot:01}.  We define the number of times $P_{n_{\alpha}}^{r}$ when $n_{\alpha}$ particles are detected on vertex $\alpha$ at step $r$ as:
\vspace*{-0.2cm}
\begin{equation}
\label{11}
P_{n_{\alpha}}^{r}=\sum_{j}\frac{(C_{j\ell}^{r})^*C_{j\ell}^{r}}{[\mathcal{K}_{r}]^2M^{2}}\frac{D\sml N-n_{\alpha},M^{2}-1\smr}{D\sml N,M^{2}\smr}.
\end{equation}
In Figs. ~\ref{fig:gridgr} we present the results obtained during the simulations of five bosons and five fermions walking on lattices {I} and {II}.  The algorithm of the computer simulations is presented in the Appendix. These vertices counting statistics describe the counting obtained for each vertex of the lattices during the time steps $r=10, 50, 100$ and $200$. In order to verify this ETH-like behavior we need to verify the counting statistics for $n_{\alpha}=5$. In fact we made the counting statistics starting from $n_{\alpha}=1, 2, 3, 4$ and $5$  and we observed the same qualitative behavior. In figures, we have only presented the counting statistics observed for $n_{\alpha}=2$ because the color gradient of the plots are more distinguishable. One can notice that the counting statistics after a certain number of time steps depending on the type of lattice and the types of quantum walkers tend to a relaxation state. This is visible in Figs. ~\ref{fig:gridgr31}--\ref{fig:gridgr44}. This observation involves differences in the colors that are not very perceptible if one doesn't have a closer look. We will give more details about these results in the next section.
\section{Thermalized quantum walks}

If we consider a $M^{\2}$ product operator
\vspace*{-0.2cm}
\begin{equation}
\label{2}
\widehat{\mathcal{N}} = \widehat{\mathbf{\Uppsi}}^{\dagger}{\sml{x}_{\1}\smr}\widehat{\mathbf{\Uppsi}}{\sml{x}_{\1}\smr}\widehat{\mathbf{\Uppsi}}^{\dagger}{\sml{x}_{\2}\smr}\widehat{\mathbf{\Uppsi}}{\sml{x}_{\2}\smr}\ldots\widehat{\mathbf{\Uppsi}}^{\dagger}{\sml{x}_{\M^{\2}}\smr}\widehat{\mathbf{\Uppsi}}{\sml{x}_{\M^{\2}}\smr},
\end{equation}
then $D\sml \mbox{\small{$N,M^{\2}$}}\smr$ configuration vectors $|{\bf n}_{\ell}\rangle$ are eigenstates of $\widehat{\mathcal{N}}$. Any state (in our case it is GMP state $|\Psi_{r}\rangle$) of the $N$ particles on the $M^{\2}$ vertices lattice can be represented as a point on a $D$ dimensional manifold in the hyperspace spanned by these vectors (i.e. configurations). A discrete evolution of such constructed many-particle closed quantum system is a mapping that transforms the coordinates of one point of the manifold into another point of this manifold.   

Many-particle systems like these have been extensively studied and theirs Hamiltonians are well known. Very recent results in this domain helped in development  of experimental quantum optics in what is called quantum matter \cite{Immanuel01}. These Hamiltonian are of Bose-Hubbard type consisting of a combination of the sum of local operators and the interaction terms depending on the  topology of the lattice. In general, such systems have a certain number of conserved global observables (e.g the total energy or the total number of particle). 
However in our case the system is not given by Hamiltonian and its evolution is not unitary. Nevertheless,  the total number of particles  $N$ is the conserved quantity. In the case of Hamiltonian evolution, the study of quantum thermalization of these kinds of systems,  can be  done using local observables such as mode occupation numbers see \cite{Rigol0}. 
To study the mechanism of quantum walks thermalization we will also monitor the dynamics using local observables connected to each vertex \cite{Pierrot:01}. This approach is similar to the one used in Hamiltonian evolution when global observables do not give the insight into the relaxation process while local observables give it \cite{Lychkovskiy2013}. Moreover, the thermodynamical limit is completely inaccessible for global observables such as the total number of particles in the system $N$.

The study of thermalization in quantum system is related to two facts: the time evolution of observables and the relaxation to a thermal state. The relaxation to a thermal state  can only be observed using  a few observables and it means the apparent loss of the system dependence on its initial conditions. This fact contradicts the unitarity of quantum evolution that depends on initial conditions. Mathematically we have two different expressions on infinite time average of mean value $\bar {O}$ of observable $\hat {O}$. The one with unitary dynamics and dependence on the initial
state encoded in coefficients $a_{\ell}$ is given in Eq. ~\eqref{0.3} and it is different from the one given by the right-hand side  of Eq. ~\eqref{0.0} produced by a thermodynamical (micro canonical) ensemble  $\hat{\rho}_{eq}$ \cite{LangenPhd}. This incompatibility makes the thermalization of quantum systems hard to understand. To solve this apparent paradox different approaches were suggested \cite{Deutsch,Srednicki}. For example it is suggested that during its evolution toward a thermal state, the system still contain the information about its initial condition but in order to access this information one must measure global operators \cite{Huse:01,Huse:02}.  In such a case it is necessary to partition the system into interconnected subsystems whereas one of the subsystem is fully defined by the initial state of the system. Consequently assessing the evolution of the system to a thermal state will be equivalent to tracking its relaxation of the subsystem defined by the initial state to a thermal state.
Many author have extensively worked on the explanation of how thermalization emerges in quantum systems \cite{Rigol0,Deutsch,Srednicki}.  The general consensus is that thermalization of a quantum system is associated with its eigenstates individually. Such a mechanism is called eigenstate thermalization hypothesis (ETH). But this ETH hypothesis doesn't explain how the thermalization of the quantum system emerges. Observations made on Hamiltonian systems undergoing quantum thermalization show that the quench process plays a crucial role of deviating the evolution of the system from a unitary evolution to a non-unitary evolution. In other words, the system is perturbed in order to drive it away from its initial dynamics to a new dynamics and when the relaxation occurs, the eigenstates thermalize. One can observe that when the relaxation occurs, the dynamics become time independent. It is crucial here to notice that even though the study of quantum thermalization is done in the Hamiltonian systems, it is not a unitary evolution because of the quench mechanism.  A possible verification of  such an hypothesis can correspond with identifying one or a couple of its eigenstates and monitoring them till the system relaxes to a thermal state.  But in Hamiltonian dynamics it is impossible to prepare a system in its eigenstate under laboratory conditions and explore the ETH mechanism because the Hamiltonian dynamics is unitary. This may be the reason why ETH remains under the label of "hypothesis".

The discrete time quantum walks described in this work gives us such an analogy. More precisely, the system constituted by many-particles on a $M^{\2}$ vertices lattice offers a natural partition. We consider each vertex as a subsystem and therefore we can study the relaxation of a specific vertex subsystem because it is being coupled to the rest of the lattice as a reservoir. We also want to mention that this way of partitioning of our system is not unique, there are many other ways.  As it was observed in our previous work \cite{Pierrot:01}, during the time evolution of the entire system the dimension of the effective configurations Hilbert space $\dim(\mathcal{H}_{e})$ increases until the relaxation to a state similar to a thermal state. This Hilbert space $\mathcal{H}_{e}$ consists of configurations with non-zero amplitudes. Moreover, while $\dim(\mathcal{H}_{e})$ increases, the dimension of the Hilbert space of the rest of the system (i.e. the lattice without the one considered vertex) increases as well. But the dimension of the Hilbert space of our subsystem consisting of just one vertex remains the same.  In this case we can study the evolution of a few observables on the subsystem we are monitoring and witness their relaxation.  Such a relaxation corresponds to a global behavior we have called quantum walks thermalization. 

But the discrete time many-particle quantum walks offer us the possibility of simulating many-particle quantum walks initially starting from a configuration, and aided by its non-unitary evolution, observe its quantum walks thermalization as well as the thermalization of the rest of the configurations. We can study the quantum walks thermalization process by means of measurements of one of the local observables involved in the relaxation process of indexed observables. We monitor the time evolution of such  observable on just the vertex $\alpha$. The quantum walks thermal relaxation state will correspond to the asymptotic value of this observable reached when the dimension of effective configurations Hilbert space tends to its stationary value. In our numerical experiment, we observe that this corresponds to the asymptotic value of the entropy of the configuration associated with the observed vertex and consequently to the temperature of that state. In this work, we are presenting the results showing that quantum walks thermalization occurs by monitoring one configuration while knowing that the ensemble of configurations relaxes. We need to precise that discrete quantum walks are non-unitary process and therefore even though the system is prepared in an precise configuration it shows the dynamics because it is similar to a sampling process over the configuration Hilbert space. Our observation on the considered  four systems on lattices {I} and {II} (see Figs. ~\ref{fig:starting})  shows that all the observables associated with the quantum walks thermalization reach asymptotic values. The numerical simulations were performed with lattices {I} and {II}  where the number of vertices ranging from $M=16$ to $M=25$ and studied the quantum walks for particle numbers ranging from  $N=4$ to $N=10$ for both multimode bosons and fermions. The results presented in this paper correspond to the case with $N=5$ particles to have the direct analogy with results obtained in \cite{Rigol0}. We have effectively observed in the quantum walks thermalization a mechanism similar to the ETH mechanism. 
\vspace*{-0.5cm}
\section{Eigenstates thermalization hypothesis probed}

\begin{figure}[htb]
    \centering
       \subfigure[5 bosons on lattice {I}]
    {
        \includegraphics[width=4.2cm, height=3.6cm]{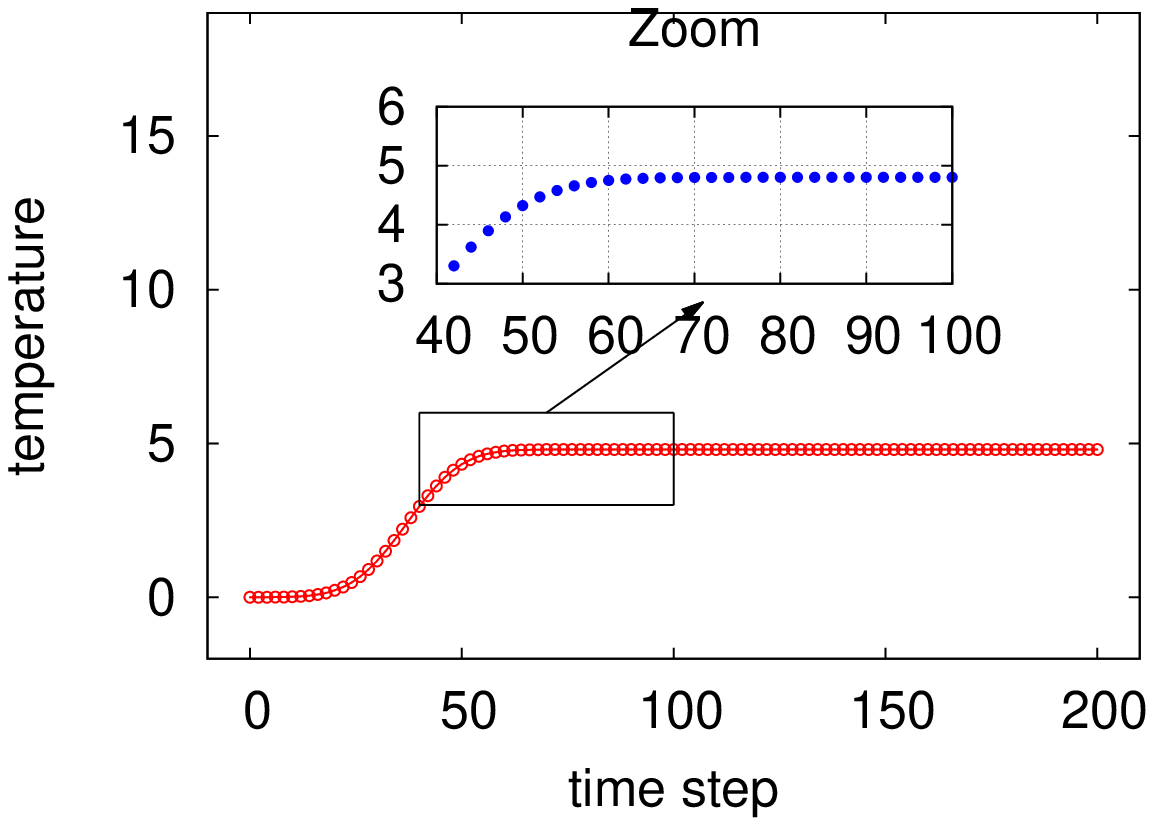}
         \label{fig:temp1}
    }
     \hspace*{-0.5cm}
    \subfigure[5 fermions on lattice {I} ]
    {
        \includegraphics[width=4.2cm, height=3.6cm]{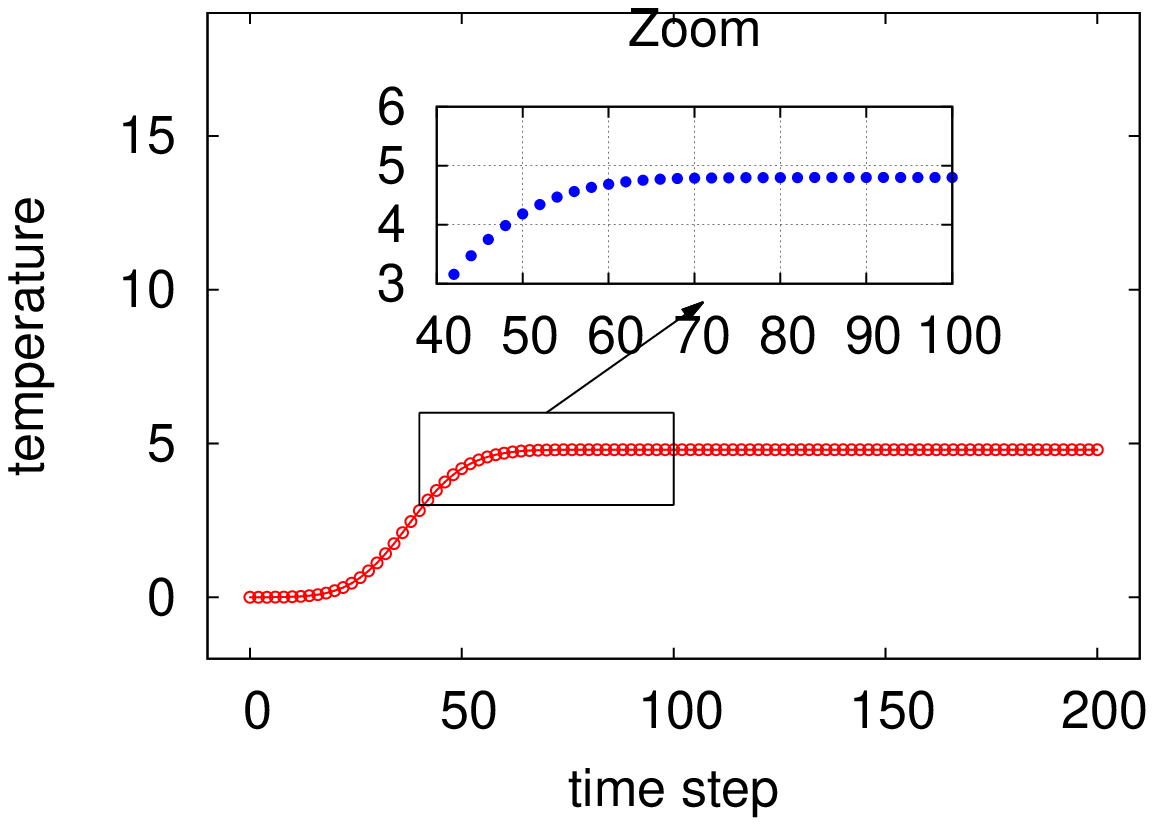}
         \label{fig:temp2}
    }\\
    \vspace*{-0.4cm}
    \subfigure[5 bosons on lattice {II}]
    {
        \includegraphics[width=4.2cm, height=3.6cm]{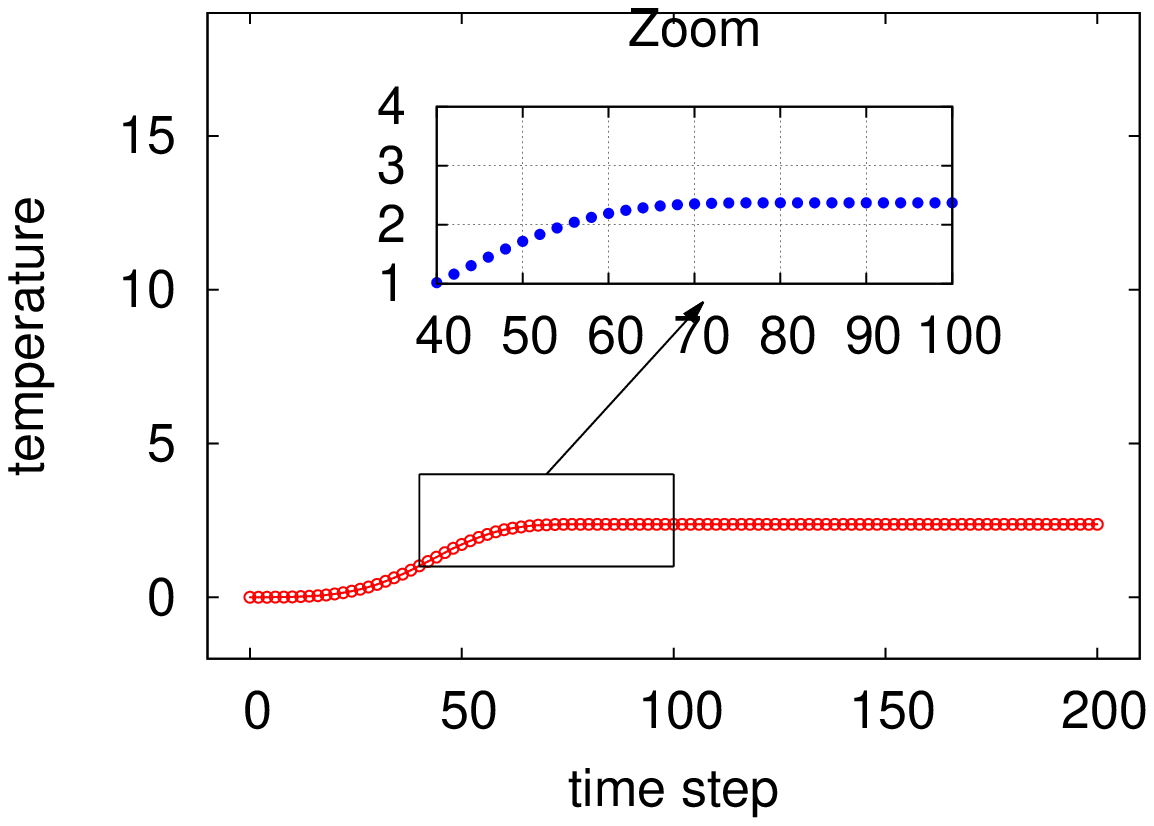}
        \label{fig:temp3}
    }
    \hspace*{-0.5cm}
     \subfigure[5 fermions on lattice {II} ]
    {
        \includegraphics[width=4.2cm, height=3.6cm]{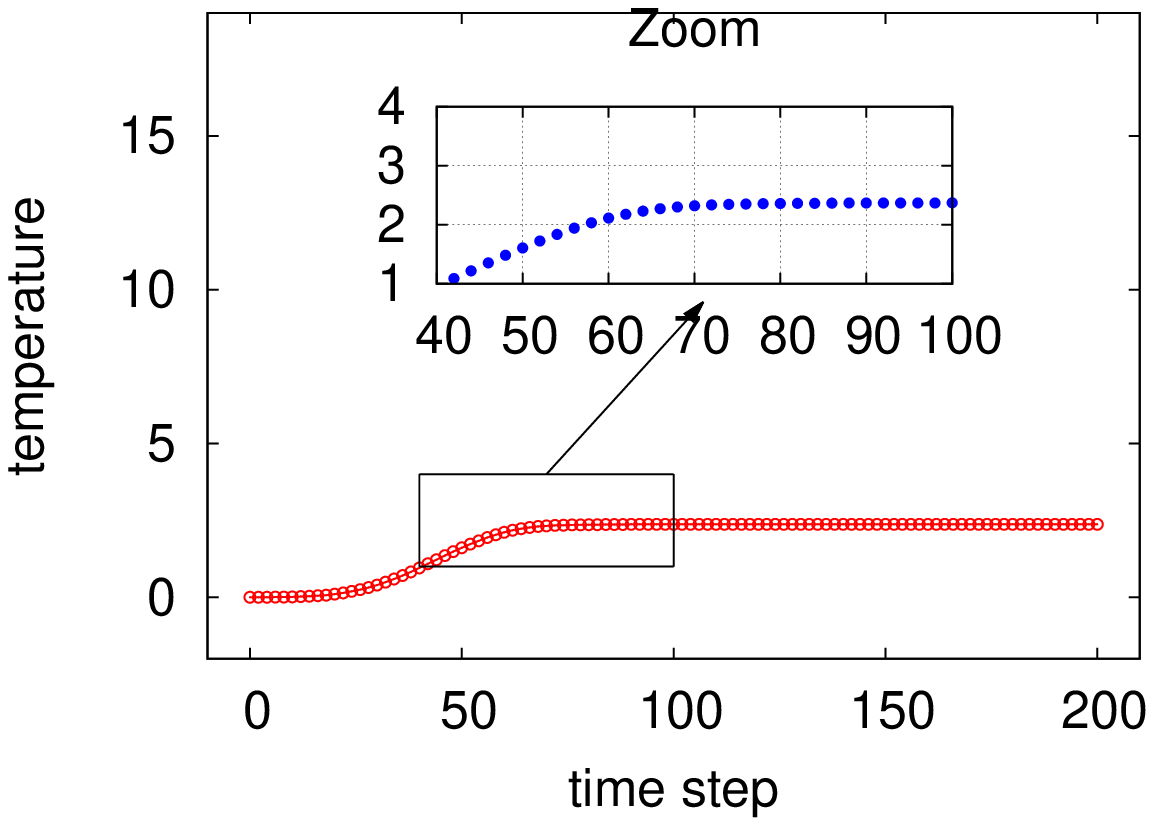}
         \label{fig:temp4}
    }
    \caption{Simulations of the temperature of the configuration $|\varphi_{r}\rangle$ given by Eq. (\ref{4.1}) for $r=200$ steps of five bosons and five fermions  quantum walks on lattices {I} and {II}.}
    \label{fig:temp}
\end{figure}

In order to effectively analyze the ETH-like mechanism in many-particle quantum walks, we have to monitor the dynamics of the two main observables: the dimension of the effective Hilbert space and the counting statistics. We used them already in \cite{Pierrot:01} and the interested reader can find in this article more details about properties and implementations of these observables to quantum walks.  The evolution of the dimension of the effective configurations Hilbert space $\dim(\mathcal{H}_{e})$ gives us a global observable for the system. We can use its time evolution as a probe at macroscopic level similarly to the global temperature that gives the macroscopic characterization of the thermalization in generic systems. In addition, we use the vertex counting statistics associated with an initial configuration from which the quantum walks were initiated. In other words, the initial state \eqref{4} is a configuration with all the particles localized on vertex $\alpha=1$. Therefore the counting statistics of $n_{\1}=N$ quantum walkers on vertex $\alpha=1$ provides the evidence that the initial configuration is a component of the GMP state during all its time evolution. In addition to the counting statistics and the dimension of the effective configurations Hilbert space, we also study the time evolution of quantum entropy of this configuration and we evaluate its temperature. The last observable gives us understanding of the evolution of this eigenstate from the first step to the last one.  Let us consider the configuration
\vspace*{-0.2cm}
\begin{equation}
 \label{4.1}
|\varphi_{\0}\rangle=\sum_{k}\frac{C_{k\ell_{\0}}^{\0}}{\mathcal{K}_{\0}}|v_{k},\,{\bf n}_{\ell_{\0}}\rangle.
\end{equation}
Suppose that during the many-particle quantum walks steps implementation the system evolves from $|\Psi_{\0}\rangle$ to $|\Psi_{r}\rangle$. More precisely when we are speaking about the configuration $|\varphi_{\0}\rangle$ (or $|\varphi_{r}\rangle$) we are referring to the configuration $|{\bf n}_{\ell_{\0}}\rangle$ without paying attention to its amplitude.  During this evolution we can record amplitude $\frac{C_{k\ell_{\0}}^{r}}{\mathcal{K}_{r}}$ of eigenstate $|\varphi_{r}\rangle$ in Eq. \eqref{4.1}  at each time step $r$ 
\vspace*{-0.2cm}
\begin{equation}
 \label{4.1.1}
|\varphi_{r}\rangle=\sum_{k}\frac{C_{k\ell_{\0}}^{r}}{\mathcal{K}_{r}}|v_{k},\,{\bf n}_{\ell_{\0}}\rangle.
\end{equation}
This configuration $|\varphi_{r}\rangle$ enters into the formation of the GMP state $|\Psi_{r}\rangle$.  We calculate the von Neumann entropy \cite{NielsenChuang} for the configuration $|\varphi_{r}\rangle$ at an arbitrary time step $r$, defined as
\vspace*{-0.2cm}
\begin{equation}
\label{4.2}
S_{\ell_{\0}r}=-k_{\B}\bigg|\frac{C_{k\ell_{\0}}^{r}}{\mathcal{K}_{r}}\bigg|^{\2}\log\bigg|\frac{C_{k\ell_{\0}}^{r}}{\mathcal{K}_{r}}\bigg|^{\2},
\end{equation}
where $k_{\B}$ is the Boltzmann constant and $\log$ is the natural logarithm. In this work we choose units in such a way that $k_{\B}=1$.  In addition,  we know that the energy in the system is provided only by the movement of quantum walkers and in our previous work \cite{Pierrot:01} we obtained the following expression for the energy per mode 
\vspace*{-0.2cm}
\begin{eqnarray}
 \label{4.3}
 \langle{E}_{\upeta}\sml r\smr\rangle&=&\frac{1}{2}\langle\varphi_{r}|\hat{\bf p}_{\upeta}^{\dagger}\hat{\bf p}_{\upeta}|\varphi_{r}\rangle\\
 \vspace*{-0.3cm}
 &=&\sum_{\ell}\sum_{j}\bigg|\frac{C_{j\ell}^{r}}{\mathcal{K}_{r}}\bigg|^{2}\frac{{n}_{\alpha}}{N}\sum_{{n}_{\upeta}}\bigg({n}_{\upeta}+\frac{1}{2}-\cos[2{\varphi}_{\upeta}{x}_{\alpha}]\bigg),\nonumber
 \end{eqnarray}
 where $\upeta$ is the mode and the mass of each quantum walker is equal to the unity. We define the total energy in the eigenstate $|\varphi_{r}\rangle$ at any given time step $r$ as the sum of the energies over all modes. Therefore the temperature of the configuration is obtained as 
 \vspace*{-0.2cm}
 \begin{equation}
 \label{4.4}
T_{\ell_{\0}r}=\sum_{\upeta}\frac{\langle{E}_{\upeta}\sml r\smr\rangle}{S_{\ell_{\0}r}}.
\end{equation}
In the Figs. ~\ref{fig:temp} we present the results of simulations of the temperature for the configuration $|\varphi_{\0}\rangle$ defined by Eq. ~\eqref{4.1} during $r=200$ steps for: five bosons on lattice {I} in Fig. ~\ref{fig:temp1}, five bosons on lattice {II}  in Fig. ~\ref{fig:temp2}, for  five fermions on lattice {I} in Fig. ~\ref{fig:temp3} and five fermions on lattice {II} in Fig. ~\ref{fig:temp4}.  In addition, we observed that the quantum walks thermalization appeared simultaneously with reaching the limit value of the dimension of the effective configurations Hilbert space $\mathcal{H}_e$ and reaching the relaxation state in the counting statistics (see Figs.\ref{fig:gridgr31}--\ref{fig:gridgr44}). 
\vspace*{-0.1cm}
More specifically, for the systems of five bosons or fermions on lattice {I} or {II}, see Fig. ~\ref{fig:starting}, we observed that after reaching a specific dimension of the effective Hilbert space $\dim(\mathcal{H}_e)$ at a certain time step $r$, each system relaxes to a thermal state and for the configuration $|{\bf n}_{\ell_{\0}}\rangle$ temperature reaches  asymptotic value of $T_{\ell_{\0}r}$ units.  We have observed:  $\dim(\mathcal{H}_e)=43402$ at time step $r=82$ and $T_{\ell_{\0}r}=4.8034$ for five bosons on lattice {I}. For the system of five bosons on lattice {II}, used in \cite{Rigol0}, we got: $\dim(\mathcal{H}_e)=21252$ at time step $r=78$ and $T_{\ell_{\0}r}=2.3715$. For systems of five fermions on lattice {I} and lattice {II} we got: $\dim(\mathcal{H}_e)=43371$ at time step $r=84$ and $T_{\ell_{\0}r}=4.8003$,  and $\dim(\mathcal{H}_e)=21245$ at time step $r=106$ and $T_{\ell_{\0}r}=2.3709$, respectively. The Fig. ~\ref{fig:temp4} shows the stabilization of the temperature of the configuration $|\varphi_{r}\rangle$ during its time evolution when the temperature reaches the value $T_{\ell_{\0}r}=2.3709$ units of temperature. These simulations presented in Figs. ~\ref{fig:temp}  show the same behavior for both bosons and fermions on lattices I and II. In some cases we continued the simulations untill $r=400$ steps but we did not notice any new phenomena and the plots are the same as for $r=200$ steps. Qualitatively the same behavior of the counting statistics, the temperature and the size of the effective configurations Hilbert space were observed for quantum walks with different number of quantum walkers and different lattice sizes. These simulations show the thermalization of the configuration $|\varphi_{r}\rangle$ for bosons and fermions on both lattices.

\begin{figure}[htb]
    \centering
       \subfigure[5 bosons on lattices]
    {
        \includegraphics[width=4.2cm, height=3.6cm]{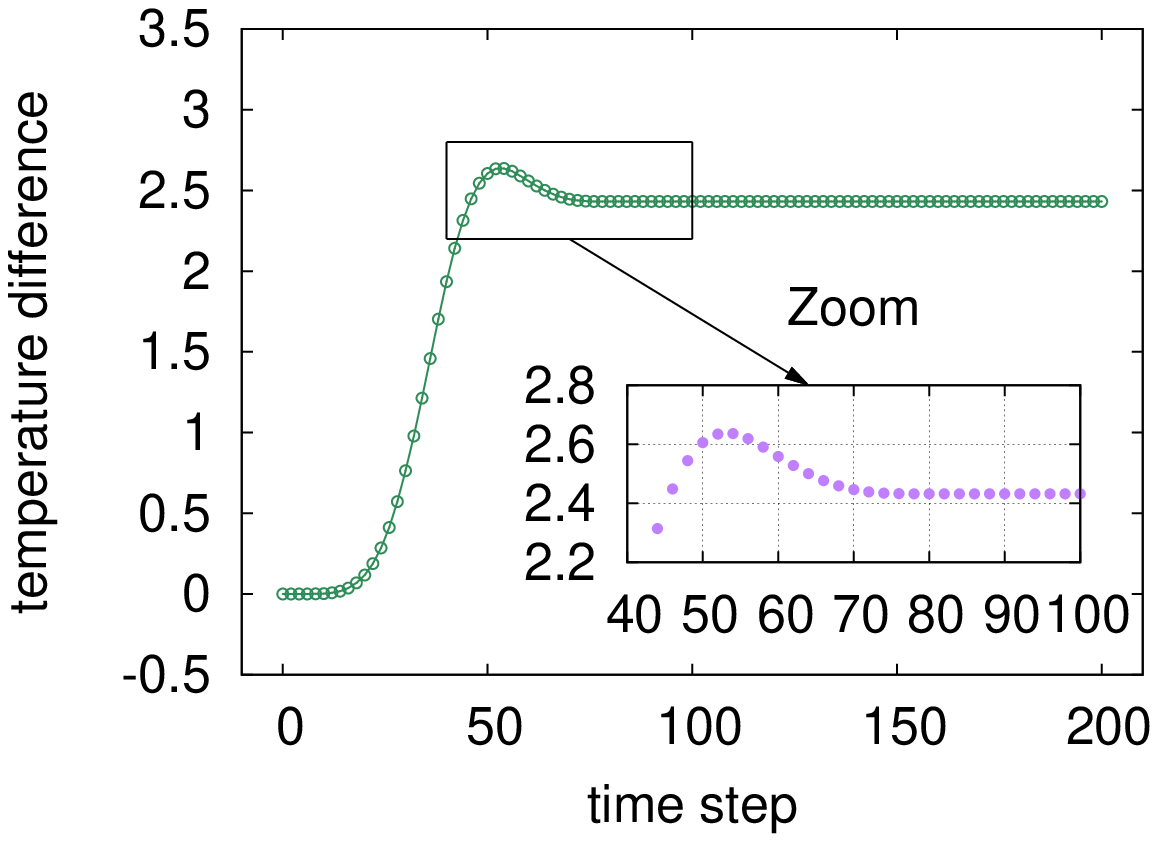}
         \label{fig:delta1}
    }
    \hspace*{-0.5cm}
    \subfigure[5 fermions on lattices]
    {
        \includegraphics[width=4.2cm, height=3.6cm]{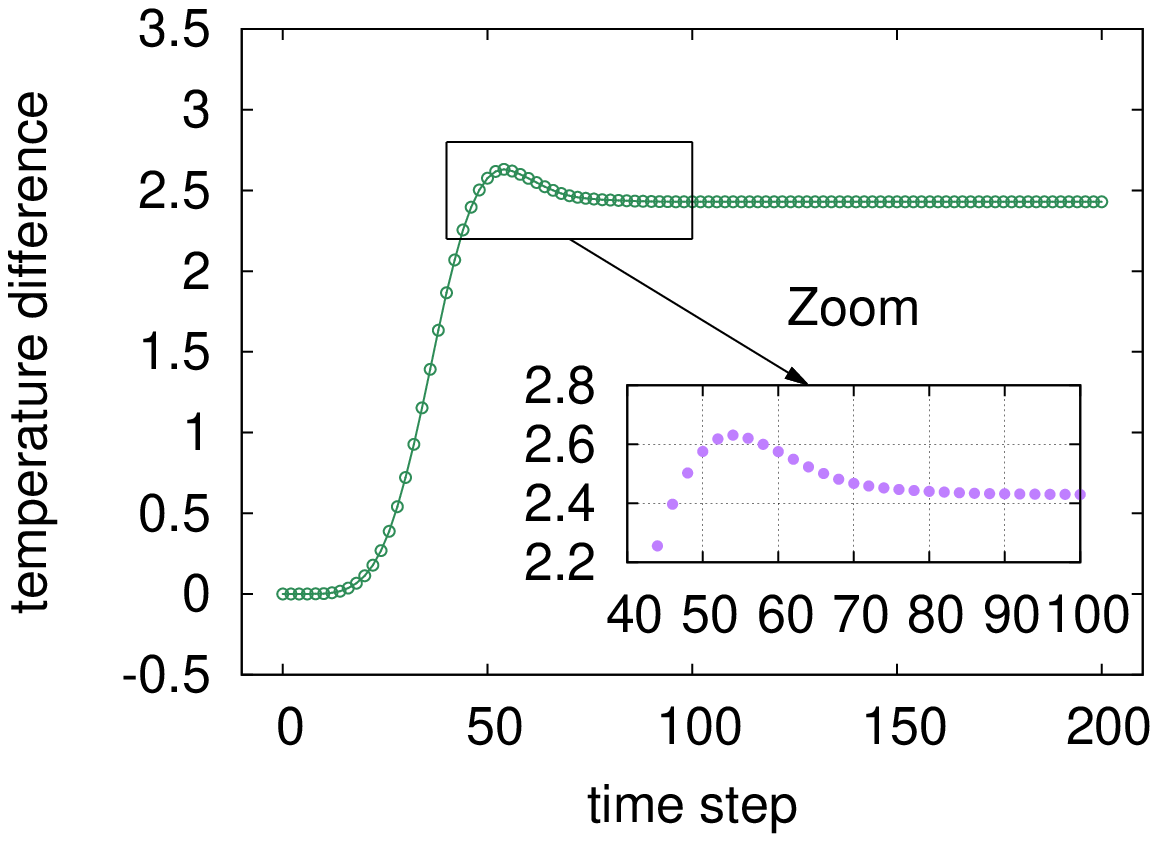}
         \label{fig:delta2}
    }
     \caption{Temperature differences between bosons systems on lattices I and II and fermions systems on lattices I and II   for $r=200$ steps of five bosons and five fermions  quantum walks.}
    \label{fig:delta}
\end{figure}
This study was conducted in comparison with the results obtained on the thermalization in generic isolated quantum systems \cite{Rigol0}. The lattice used for that study is what we call lattice {II}  (see Fig.~\ref{fig:starting4}) and the number of particles was also 5. Our results shows that the quantum walks thermalization doesn't depend on the topology of the system under study. This is established by our calculations for a fully connected lattice called lattice I with the same number of particles that gave the same qualitative results as for lattice II. The only difference in dynamics on both lattices appears in the times taken for many-particle quantum walks systems to reach thermalization. For the fully connected lattice I the particles spread without obstacles and the time is shorter while for lattice II the flow between the two sub-lattices is slowed because they are connected by just one edge.
\section{Conclusions and outlooks}

An additional comparative observation can be made on the quantum walks thermalizing systems on the basis of the system{-}reservoir division made earlier. In fact, we observe that the exchange system{-}reservoir here happens through the degree of connectivity between the two systems in both cases concerning bosons and fermions. We compare the temperature differences between systems of five bosons  (see Fig. ~\ref{fig:delta1})  and five fermions (see Fig. ~\ref{fig:delta2}).  We observe that larger systems (with larger number of edges between vertices) have higher temperature, see Fig. ~\ref{fig:temp1} for bosons and  Fig. ~\ref{fig:temp3} for fermions in comparison with the temperature of smaller systems, see Fig. ~\ref{fig:temp2} for bosons and Fig. ~\ref{fig:temp4} for fermions, respectively. This can be explained by the fact, that in the larger system with more edges between vertices given in Fig. ~\ref{fig:starting3} the GMP state splits into more components and the dynamics takes place in a bigger subset of the configurations Hilbert space.  For the system with a small number of edges between vertices given in Fig. ~\ref{fig:starting4}, the GMP state is for a finite time defined by a finite number of components.  Therefore the dynamics for the systems in lattice {I} is driven by the rapid expansion of the effective configurations Hilbert space while for the systems in lattice {II} the dynamics is slowed by restricted connectivity between vertices that also slows the expansion of the effective configurations Hilbert space.

Even though in this paper we only present the results for the case of $N=5$ quantum walkers on two $M^{\2}=25$ vertices lattices (see Fig.~\ref{fig:starting}) we checked the universality of observed thermalization-like phenomena using different numbers of walkers $N$ and number of vertices $M^{\2}$ of square lattices. Namely, we made quantum walks simulations with $N=4, 6, 8, 10$ particles on lattices I and II with $M^{\2}= 25$ vertices. We also made calculations on two lattices with $M^{\2}=16$ vertices for $N=4, 6, 8,10$ particles analogous to lattice I and II, namely fully connected lattice and lattice consisting of two sub-lattices with 4 and 10 vertices respectively. The results obtained were qualitatively similar. The only difference observed was in the time taken for the many-particle quantum system to thermalize. The thermalization time increases with the number of particles for the same lattice.  In this paper we've only presented the results for 5 particles in comparison to the results obtained in \cite{Rigol0} concerned thermalization in Hamiltonian system.  One can also notice that thermalization times are shorter for lattices with smaller number of vertices with the same topology.

Our study shows that many-particle quantum walks on lattices exhibit an ETH-like behavior and can therefore be used to probe the ETH effectively. In this study we used a simple configuration $|\varphi_{r}\rangle$ with all particles at one vertex as the eigenstate to perform the probe on ETH.  As we mentioned, we partitioned the lattices in vertex subsystems. We can also partition systems using groups of vertices with a preferred choice of occupation numbers on that group of vertices. In the same way,  we can also use every single configuration in the configurations Hilbert space and obtain qualitatively the same behavior of observables. Moreover this study also shows that the quantum walks thermalization of the many-particle quantum walks system doesn't destroy the coupling between the system and its initial condition. The thermalized configuration remains present in the effective Hilbert space of the system during its time evolution toward a quantum walks thermalized state similarly to what was announced in \cite{Huse:01,Huse:02}. The continuity of the temperature graph obtained in our simulations is the proof that the initial configuration was present at every time step. In addition, the dynamics of these systems on lattices confirm our previous conjecture \cite{Pierrot:01} that any quantum walks thermalization should correspond to a relaxation of the dimension of the effective configurations Hilbert space. 

The challenge of such studies remains the level of calculations complexity quickly growing with the size of the lattice and the number of particles which translates into a rapid growth of the dimension of the effective configurations Hilbert space that consists of configurations with non-zero amplitudes. The step by step implementation is similar to exploring the entire configuration Hilbert space and at each step assigning non zero amplitudes to an increasing number of configurations. As the result, the thermalization time grows with the size and the number of particles. In the worst case, each configuration can generate 80 new configurations in one time step. Each obtained configuration must be compared to the existing ones before being added to the list or merged to the existing list. Thus all configurations from previous steps must be kept in the register. As the result, after $r$ steps the memory contains $80^{\2r}$ registers. This computation operations are highly memory demanding.  We used C++ and a short algorithm is presented in the Appendix.  In addition, it will be interesting to consider quantum walks in cases where the number of particles exceed the number of vertices.  The complexity of such computer simulations is time and resources consuming and need a different approach which will be presented in the future paper.  These models are also interesting because they can be realized in laboratories of quantum optics and quantum engineering where arrays of coupled cavities and trapped atoms are used in some experiments. We can notice that computer simulations give us more opportunity to explore evolution of an initial state with a large number of configurations in contrary to the limitation of the laboratories conditions. 

This research was supported by grant No. DEC-2011/ 02/A/ST1/00208 of National Science Centre of Poland.

\section*{Appendix: Algorithm of the computer simulations}
We use the case of 25 vertices two dimensional lattice and one initial configuration.
\subsection*{Module 1: Initialization}
This module consists of preparing the initial configuration and its amplitudes. 
\begin{enumerate}
\item Define an array of the dimensions of the lattice (the initial configuration).
The defined array should be initialized such that all the cells are empty; cells are vertices here labelled by $C$.
\item Populate the occupied vertices (In our case cell 0 had to receive 5 particles).
\item Define a 4 dimensional vector (amplitude).
Cells are sides of the coins. In our case cell 2 and 3 must be non-zero and normalize.
\end{enumerate}
\subsection*{Module 2: Quantum walks implementation}
This module combines the coins operation and the particle shifting in order to generate the quantum walks dynamics. It results into generating a list of configurations using the initial configuration. For the 25 vertices square lattice each cell labelled C is surrounded by four cells respectively labelled $C-1,C+1,C-5$ and $C+5$.   
\begin{enumerate}
\item Select a configuration.
\item Duplicate the selected configuration
\item Identify the first occupied vertex (cell $C$ with non-zero value). 
\item Check amplitude cells one by one. 
\item If the first is non-zero, add one to Cell $C-1$ of the duplicated configuration,
\item If the second is non-zero add one to cell $C+1$ of the duplicated configuration,
\item If the third is non-zero, add one to Cell $C-5$ of the duplicated configuration,
\item If the fourth is non-zero add one to cell $C+5$ of the duplicated configuration,
\item Remove one from Cell $C$ of the duplicated configuration.
\item Save the duplicated configuration.
This adding and removing are quantum particles creation and annihilation operations  therefore the residuals factors should be used in the calculation of the amplitudes of the duplicated configurations.
\item Compute its amplitudes using Eq. \eqref{9} 
\item Move to the next non-zero cell of the configuration in 1, repeat the process 2--11 on all its non-zero cells. 
\item Normalize the amplitudes
\end{enumerate}
\subsection*{Module 3: Measurements}
The aim of this module is to compute the expected values of the observables that later are plotted.
At the end of module 2 the code generates two registers: the amplitudes and the configurations. We use these register to:
\begin{enumerate}
\item compute the vertices expectation numbers,
\item compute the entropy,
\item compute the temperatures,
\item compute the counting statistics of a specific number of particles on each vertex for configuration for a targeted number of particles using Eq. \eqref{11}. 
\end{enumerate}
Repeat this process on all existing configurations as many times as the intended  number of steps.

This algorithm can be modified and adapted to different sizes of square lattice. In addition, one need to the adjacency matrix in the test process to monitor the connectivity for the boundary vertices and the corner ones.


\end{document}